\documentclass[preprint, superscriptaddress, a4paper, 
prb]{revtex4}
\usepackage[utf8]{inputenc}
\usepackage[paperwidth=210mm,paperheight=297mm,hmargin=2cm,vmargin=1.5cm]{geometry}

\usepackage{amsmath}
\usepackage{amsfonts}
\usepackage{amssymb}
\usepackage{graphicx}
\usepackage[margin=10pt,font={scriptsize, singlespacing},labelfont=bf]{caption}
\usepackage{subfig}
\usepackage[T1]{fontenc}
\usepackage{ae}
\usepackage{hyperref}

\usepackage{overpic}

\graphicspath{{graphics/}}

\usepackage{xcolor}

\newcommand{\zweivector}[2]{\left(\begin{matrix} #1 \\ #2 \end{matrix}\right)}

\newcommand{\absq}[1]{\left\vert #1 \right\vert^2}
\newcommand{\abs}[1]{\left\vert #1 \right\vert}
\newcommand{\argument}[1]{\mathrm{arg}\left( #1 \right)}

\begin{document}
\title{Accessing topological superconductivity via a combined STM and renormalization group analysis}
\author{Lars Elster}
\affiliation{Institute for Theoretical Physics, TP IV, University of W$\ddot{u}$rzburg, Am Hubland, D-97074 W$\ddot{u}$rzburg, Germany}
\affiliation{Univ. Grenoble Alpes, INAC-SPSMS, F-38000 Grenoble, France \\ CEA, INAC-SPSMS, F-38000 Grenoble, France}
\author{Christian Platt}
\author{Ronny Thomale}
\author{Werner Hanke}
\affiliation{Institute for Theoretical Physics, TP I, University of W$\ddot{u}$rzburg, Am Hubland, D-97074 W$\ddot{u}$rzburg, Germany}
\author{Ewelina M. Hankiewicz}
\email[]{ewelina.hankiewicz@physik.uni-wuerzburg.de}
\affiliation{Institute for Theoretical Physics, TP IV, University of W$\ddot{u}$rzburg, Am Hubland, D-97074 W$\ddot{u}$rzburg, Germany}

\date{\today}

\begin{abstract}
The search for topological superconductors has recently become a key issue in condensed matter physics, because of their possible relevance to provide a platform for Majorana bound states, non-Abelian statistics, and fault-tolerant quantum computing. We propose a new scheme which links as directly as possible the experimental search to a material-based microscopic theory for  topological superconductivity. For this, the analysis of scanning tunneling microscopy, which typically uses a phenomenological ansatz for the superconductor gap functions, is elevated to a theory, where a multi-orbital functional renormalization group analysis allows for an unbiased microscopic determination of the material-dependent pairing potentials. 
The combined approach is highlighted for paradigmatic hexagonal systems, such as doped graphene and water-intercalated sodium cobaltates, where lattice symmetry and electronic correlations yield a propensity for a chiral singlet topological superconductor state.
We demonstrate that our microscopic material-oriented procedure is necessary to uniquely resolve a topological superconductor state.
\end{abstract}

\maketitle

\section{Introduction}

The search for topological states of matter has recently generated a flurry of broad activity in the field of superconductivity (for several paradigmatic directions, see for example Refs.~\onlinecite{sigrist_1991,Read_2000,mourik_2012, das_2012, Maier_2012, fu_2008, qi_2009,PhysRevLett.105.097001,qi_2011,tkachov_2013,tkachov_2013b}). A topological superconductor (SC) is an unprecedented state of quantum matter, which possesses a full pairing gap in the bulk but gapless exotic surface states if a surface termination exists, as well as possibly non-trivial vortex bound states \cite{Read_2000,PhysRevLett.86.268,fu_2008}. A chiral superconductor \cite{Jackiw_1976,sigrist_1991,Read_2000} with broken time-reversal symmetry (TRS) may be considered the superconducting analogue of the quantum Hall phase (as characterized by a non-trivial Chern number of its Bogoliubov bands~\cite{thouless_1982}), whereas a topological superconductor conserving TRS \cite {qi_2009} is closely related to the quantum spin Hall phase (along with its non-trivial $\mathbb{Z}_2$ invariant \cite{kane_2005}). Recently, chiral SCs have enjoyed significant attention, exhibiting a variety of exotic phenomena based on their non-trivial topology \cite{sigrist_1991}, such as hosting Majorana vortex bound states \cite{Read_2000,PhysRevLett.86.268} and gapless chiral edge modes, that carry quantized thermal or spin currents (see for example Ref.~\onlinecite{Horovitz_2003}). The Majorana bound states can be interpreted as elusive fermionic particles equivalent to their own antiparticles, and have potential applications in fault-tolerant topological quantum computation \cite{Nayak_2008}.

In view of these striking properties, it would be desirable to develop guidelines to identify materials with the potential to host chiral topological SC states. As it turns out, the interplay of the lattice symmetry, the shape and the size of the Fermi surface (fermiology), the multi-orbital character, and the electron-electron (e-e) interaction are decisive for an unconventional chiral pairing mechanism.

Material-specific research into this direction first concentrated on the perovskite Sr$_2$RuO$_4$, where experimental evidence points to a chiral odd-parity p-wave SC state \cite{mackenzie_2003}, as a possible analogue of superfluid $^3$He \cite{volovik_1988}. However, the topologically protected Majorana edge modes, which should appear in the chiral p-wave superconductor when a half-quantum vortex is injected \cite{Kallin_2012}, have – so far – not uniquely been identified, despite strong experimental efforts \cite{kashiwaya_2011}. It suggests that the odd-parity pairing, along with its non-trivial spin dependence, induces challenges which in terms of complexity even overshadow the original task to identify a material with topological chiral SC.

Therefore, in this work, we focus the attention on the combined experimental and theoretical verification of topological SC and consider even-parity topological chiral SC. This happens for a d+id topological state (with the Chern number for the Bogoliubov bands $C=\pm 2$ \cite{PhysRevB.78.195125,kitaev-periodic,neupert_2014}).
While it is, in principle, possible to obtain Majorana edge modes in the d+id SC state in the presence of both Rashba spin-orbit coupling and a weak Zeeman field \cite{PhysRevB.82.134521, blackschaffer_2012, blackschaffer_2014}, we consider here systems without spin-orbit interactions, a simplification which does not prevent us from accurately tracking down promising regimes for a d+id SC state.

On a square lattice, where most unconventional superconductors are found, the difficulty in realizing such a d+id singlet state is that the generic fermiology and interactions, which can yield a d-wave state, favour d$_{\mathrm{x}^2-\mathrm{y}^2}$ over d$_{\mathrm{xy}}$ pairing. This changes for hexagonal lattices, where the lattice symmetry protects the degeneracy of the d$_{\mathrm{x}^2-\mathrm{y}^2}$-wave and the d$_{\mathrm{xy}}$-wave SC at the instability level. This then yields a chiral singlet d$_{\mathrm{x}^2-\mathrm{y}^2}$ + i d$_{\mathrm{xy}}$  - superconducting state below the critical temperature, T$_{\mathrm{C}}$, to maximize the condensation energy \cite{blackschaffer_2014,platt_2014}.

In contrast to p+ip odd-parity pairing, the singlet character of the unconventional pairing should make its emergence more generic, as it stems from electron-mediated pairing where large wave vector particle-hole fluctuations tend to drive singlet superconductivity. This has been recently addressed in several theoretical scenarios of a d+id state such as for doped graphene \cite{blackschaffer_2007,PhysRevB.78.205431,PhysRevLett.100.146404,pathak_2010,nandkishore_2012,kiesel_2012,PhysRevB.85.035414} (for a recent review see Ref.~\onlinecite{blackschaffer_2014}), 
water-intercalated sodium cobaltates \cite{kiesel_2013}, and the pnictide SC SrPtAs \cite{fischer_2013}. ARPES data on chemically-doped graphene show that the highly-doped regime close to the van Hove singularity, albeit with a currently high degree of disorder, is  accessible \cite{mcchesney_2010}. Still, superconductivity in graphene has not yet been experimentally confirmed. From this perspective, water-intercalated sodium cobaltates and the pnictide SrPtAs may be more promising because, in both of these compounds, superconductivity has  been  already discovered \cite{takada_2003,doi:10.1143/JPSJ.80.055002}. Furthermore, some indications of unconventional SC were observed in Knight shift data on cobaltates \cite{zheng_2006}, as well as in muon-spin rotation/relaxation measurements and nuclear quadrupole experiments on SrPtAs \cite{biswas_2013,brueckner_2013}. While no unambiguous experimental confirmation of chiral d-wave SC for these materials exists so far, we believe that further experimental attention is certainly warranted and could lead to the first unambiguous identification of chiral topological SC.

 A characteristic challenge in the search for unconventional chiral SC is the pronounced competition between different orders, such as spin density-wave (SDW) and different SC orders, in particular the TRS-broken d+id SC state and an f-wave  state with TRS \cite{platt_2014,PhysRevB.89.144501} on the hexagonal lattice. This clearly calls for methods which are capable of distinguishing the competing channels at the instability level, this means at low energies of the order of k$_\mathrm{B}\mathrm{T}_\mathrm{C}$ of the SC gap features. This is the main strength 
of the functional renormalization group (fRG) method (for reviews see for example Refs.~\onlinecite{RevModPhys.84.299, platt_2014}). The fRG allows for a systematic connection via renormalization between a high-energy bare Hamiltonian and a low-energy k$_\mathrm{B}\mathrm{T}_\mathrm{C}$ effective theory, where the SC channels can be resolved \cite{platt_2014}.

Knowing the precise functional form of the pairing function from the fRG calculations is of fundamental importance to make contact with experimental signatures. 
Therefore, we combine the microscopic theory, which is the fRG method, with the theory of scanning tunneling spectroscopy (STM). Although the main task of our work is to elevate spectroscopic signatures to provide evidence in favour of a possible singlet chiral topological SC state, for the role-model systems graphene and cobaltates, this approach can be straightforwardly extended to other classes of topological SC.

The conventional STM theory has been successful in a variety of situations, where it has been viewed as a phenomenology, assuming a certain ansatz for an order parameter \cite{kashiwaya_1996,kashiwaya_2000, yamashiro_1997}. Here we demonstrate, that our microscopic material-oriented procedure is necessary in the case of competing anisotropic SC channels such as d+id, and f-wave:  the phenomenological approach (including only a single harmonic) yields qualitatively different spectra from the the full microscopic fRG + STM results, also ruling out the possibility to distinguish gapped and nodal SC order parameters on the basis of the out-of-plane STM signal alone.

\section{Results}
\subsection*{Combined fRG and STM Method for Graphene and Cobaltates}

A graphene monolayer and water-intercalated cobaltates are considered as proto-typical examples where, as shown in Fig. \ref{fig:phasescheme}, the fRG method predicts the possibility of a chiral d+id SC state. In this manuscript, we propose the combined fRG + STM method as a powerful tool to distinguish between 
d+id and f-wave order parameters in these materials.

Starting with the Hamiltonian for the cobaltates (see Sect. IV) which only includes on-site electron-electron interactions, the fRG flow will generate longer range electron-electron interactions.
Since the longer range electron-electron interactions cause higher order harmonics to be induced in the d+id-wave SC channel,  the in-plane STM signal will show qualitatively new features in comparison with the phenomenological d+id SC channel, as presented in
Figs.~\ref{fig:graphene_short}a and \ref{fig:cobaltates}a, for graphene and water-intercalated cobaltates, respectively. 
This immediately shows  the necessity to apply the fRG + STM method. Furthermore, as shown in Figs. \ref{fig:graphene_short}, \ref{fig:cobaltates} and \ref{fig:out}, the fRG + STM approach gives a new methodology, which allows for a reliable distinction between the topological d+id and the f-wave SC phase.
To fully appreciate the strength of the fRG + STM approach, we first explain the advantages of the fRG method and how the fRG input is fed into the material-oriented STM calculations.

In the cobaltates, the fRG starts from a 3-orbital model (see Sect. IV), where a core ingredient is the effect of longer-range hoppings, shifting the filling of perfect nesting away from the van-Hove filling. 
The $C_{6\mathrm{v}}$ acts as the decisive symmetry which can yield a chiral d-wave below the critical temperature $\mathrm{T}_\mathrm{C}$ to maximize the condensation energy. 

 Several common features as compared to the cobaltate case can be identified in graphene, such as the role of longer-range hopping in providing a distinction between van-Hove filling and the filling of perfect nesting. 
In particular, we discuss the chiral d-wave state, which competes with f-wave SC further away from van-Hove filling and turns into a SDW state near van-Hove filling beyond a certain interaction strength. 
Of all theoretical approaches, mainly the fRG has been capable of fully describing such a scenario \cite{kiesel_2012,PhysRevB.85.035414}.

More precisely, in graphene, we illustrate the fRG + STM analysis of shorter- and longer-range Hubbard interactions on a generalized honeycomb tight-binding model up to third-nearest neighbour hybridization (Sect. IV).  
As seen in Fig.~1a, at the van-Hove singularity vHs (orange area), chiral d+id pairing competes with, but wins over the spin-density wave channel (details for the underlying band structure and the Hamiltonian are summarized in the Section IV). 
Away from the vHS (blue area), the critical instability scale $\Lambda_\mathrm{C}$, $\Lambda_\mathrm{C} \sim \mathrm{T}_\mathrm{C}$ drops and whether the d+id or the competing f-wave instability is preferred depends on the range of electron-electron interactions. 
We show differential conductance plots for graphene with short-range Coulomb interactions (Fig. \ref{fig:graphene_short}) and long-range Coulomb interactions (Supplementary Figure 1 and Supplementary Discussion 1), because the carrier density and the corresponding range of 
electron-electron interactions is different due to charge screening for the two different doping situations.

At $\Lambda_\mathrm{C}$, which denotes the critical fRG flow parameter, where the leading instability starts to diverge, the different channels such as the SC d+id and f-wave channels are decomposed into different eigenmode contributions and the corresponding gap form factors are obtained (see Sect. IV). They are then used as a microscopic input into the STM procedure. The black dots in Fig. \ref{fig:phasescheme} denote the phase-diagram points for which the TRS-breaking d+id-phase and the TRS-conserving f-phase are calculated and taken as inputs in the STM scheme. For the cobaltates, the doping has been confined for both d+id and f-phases to the value $x \cong 0.3$, where SC has been observed experimentally.

 In the second step the differential conductance of these quasi-2D SC is calculated using a normal metal-insulator-anisotropic SC (N-I-S) setup with the pairing potentials from the fRG calculation. Such a N-I-S junction, formulated for a $\delta$-function barrier model \cite{kashiwaya_1996}, is known to imitate very well an experimental STM setup. This is documented by a variety of applications, where the pairing potentials have been used as a phenomenological ansatz for distinguishing between different SC symmetry channels \cite{kashiwaya_2000, kashiwaya_1996, yamashiro_1997, kashiwaya_2011}. The effective barrier height is represented by $Z_0=\frac{mH}{\hbar^2 k_{\mathrm{FN}}}$, where $m$, $k_{\mathrm{FN}}$ and $H$ denote the electron mass, the Fermi momentum on the normal side, and the barrier potential, respectively. Both in-plane and out-of-plane setups are considered, where in the first (second) case the STM lies in (is oriented perpendicular to) the SC plane. Solving, as in previous phenomenological studies, the Bogoliubov-de Gennes (BdG) equations with the appropriate boundary conditions, but now with the momentum dependent microscopic pairing potentials, the coefficients (probabilities) for Andreev reflection $r_\mathrm{A}$ and normal reflection $r_\mathrm{N}$ are obtained. Via the Blonder-Tinkham-Klapwijk (BTK)-formula \cite{blonder_1982}, the conductance is given by
\begin{equation}
\sigma_\mathrm{S}(E,\theta)=1+\absq{r_\mathrm{A}(E,\theta)}-\absq{r_\mathrm{N}(E,\theta)}
\label{eq:btk}
\end{equation}
for the quasiparticle injection with energy $E = -\mathrm{e}V$, where $V$ is the bias voltage, and $\theta$ is the incident angle with respect to the interface (Fig. \ref{fig:angles}). More details for the calculation of the normalised differential conductance can be found in the Sect. IV. Here it suffices to note that the conductance contains two distinct pairing potentials $\Delta_+$ and $\Delta_-$. They correspond to the effective pairing potentials for transmitted electron-like quasiparticles (ELQ) and hole-like quasiparticles (HLQ), respectively, as shown in Fig. \ref{fig:angles}. The total conductance is given by integrating the angle-resolved conductance of Eq. \eqref{eq:btk} over all transverse momenta, which are the independent modes. Except in the low-barrier (small $Z_0$-) limit, which is not of interest here (our choice of $Z_0=5$ corresponds to a high barrier), the conductance peak corresponds to the Andreev bound-state (ABS) energy level, which is formed at the edge of the superconductor (see Fig.~3 
of Ref.~\onlinecite{kashiwaya_1996}).

\subsection*{Differential conductance spectra for the in-plane setup}


In the previous sections a theory for the tunneling spectroscopy of a N-I-S junction, combined with the microscopic fRG derivation of the underlying pairing potentials, was proposed as a new and efficient tool for identifying, in particular, a chiral SC state with broken TRS.

The differential conductance for the in-plane setup is shown in polar plots (left parts of Figs. \ref{fig:graphene_short} and \ref{fig:cobaltates}), where the radial axis is the quasiparticle excitation energy normalised by the superconducting energy (gap) scale $\Delta$, which is a common energy scale obtained from fRG. $\alpha$ denotes the angle between the interface normal (n) and the k$_\mathrm{x}$-direction (see Fig.~\ref{fig:angles}). The absolute value of the pairing potential is also shown in a polar plot in order to compare it to the differential conductance. For each phase, cross sections at three different angles $\alpha$ are given in the upper right panel. The chosen angles are indicated by black lines in the dI/dV characteristics.

More specifically, the differential conductance spectra, obtained in the in-plane setup simulating an STM experiment, are presented for the d+id- and f-pairing phases of graphene (Fig.~\ref{fig:graphene_short}) and for water-intercalated sodium cobaltates (Fig.~\ref{fig:cobaltates}).
Comparing the conductance spectra (dI/dV) for the d+id-pairing phase (Fig.~\ref{fig:did_short}) with the f-pairing phase (Fig.~\ref{fig:f_short}), it becomes clear why the fRG + STM calculation is an ideal tool to distinguish a d+id-wave TRS-broken phase from a f-wave TRS conserving phase. In  graphene, for the chiral d+id case, the polar dI/dV intensity plot displays, in addition to the outer structure (which maps the local density of states with the peaks appearing at the local pairing potentials – compare with the pairing potential in Fig. (\ref{fig:did_short})) an inner structure for energies smaller than the superconducting band gap, which is called sunflower structure in what follows. It is this inner coloured sunflower structure which immediately signals the presence of a TRS-broken SC state. A similar, but even more complicated inner structure for the cobaltates (below the band gap $E=6.7 \Delta$) is again the sign of broken TRS (see Fig.~\ref{fig:cobaltates}a).

In order to understand the physical content of these differential conductance spectra for the d+id and f-pairing phases in more detail, let us summarise what kind of physical quantity the tunneling spectroscopy is detecting. The energy level giving the conductance peak is determined by a quantum condition of the bound quasiparticles (QPs) in a pseudo quantum well. This quantum well is formed by the N-I-S junction (Fig. \ref{fig:angles}), where an electron  injected from the N-side is transmitted into the superconductor (S) ejecting an Andreev hole-like quasiparticle. This hole-like quasi-particle with its wave vector k$_{\mathrm{FS}}^-$ scatters into an electron-like quasiparticle with k$_{\mathrm{FS}}^+$ after reflection from the I-S interface, with a corresponding change in the effective pair potentials from $\Delta_- = \Delta(k_{\mathrm{FS}}^-)$ to $\Delta_+ = \Delta (k_{\mathrm{FS}}^+)$ at the insulator (I) (Ref.{\onlinecite{kashiwaya_1996}, in particular Fig.~3 therein)}.  The bound states form at the interface and the tunneling electrons in the N-I-S junction flow via these bound states. These states have been shown in earlier work by Tanaka and coworkers \cite{tanaka_2005} to converge to the edge states of the superconductor (trivial or non-trivial) in the large barrier-height limit (which is considered here).

The analogy with the quasiparticles and their Andreev reflection in a pseudo quantum well is useful for a transparent physical understanding of the inner sunflower structure appearing in the STM spectra of Figs \ref{fig:did_short} and \ref{fig:did_cobaltates}, as discussed below. The quantum-well analogy has been suggested by Kashiwaya et. al. \cite{kashiwaya_1996} and shows the equivalence of the bound QP condition of the N-I-S junction with that of QPs in a S-N-S structure (with thickness of $N \rightarrow 0$ and no difference of the superconducting phase across the junction), in which the pair potentials of the two superconductors are $\Delta_+$ and $\Delta_-$, respectively. The QPs in the pseudo quantum well (normal region N) are confined if their energies are less than the amplitudes of both pair potentials, this is $E < \text{min} (\abs{\Delta_+}, \abs{\Delta_-})$. The bound QPs travel along a closed path by repeating Andreev reflections at both N-S and S-N interfaces. This is, then, equivalent to the $\Delta_+$ and $\Delta_-$ scattering processes at the I-S interface, as schematically shown in Fig. \ref{fig:angles}.

The corresponding peak condition for the bound QPs, which was first reported by Kashiwaya et.al. \cite{kashiwaya_1996}, is given in Eq. \eqref{eq:peak_condition} of Sect. IV. Here we consider its general properties, which help to understand the occurrence of the outer peaks and that of the inner sunflower structure in Figs. \ref{fig:graphene_short} and \ref{fig:cobaltates}.

For $\Delta_- = \Delta_+$, the peak condition, given in Eq. \eqref{eq:peak_condition}, is fulfilled if the energy E of the injected particle is $E = \abs{\Delta_{\pm}}$. Consequently, a peak at the bandgap occurs, which is the outer structure in Figs. \ref{fig:graphene_short}, \ref{fig:cobaltates}. We call this peak the local bandgap peak, since as shown in these figures, the bandgap value depends  on the (local) polar angle $\phi$. The polar angle is defined as $\phi = \arctan \frac{k_{\mathrm{y}}(\phi)}{k_{\mathrm{x}}(\phi)}$, with $k_{\mathrm{x}}(\phi) = k_\mathrm{F} (\phi) \cos \phi$  and $k_{\mathrm{y}}(\phi) = k_\mathrm{F} (\phi) \sin \phi $ and is given for ELQ by $\phi_+ = \theta_S - \alpha$ and for HLQ by $\phi_- = \pi - \theta_\mathrm{S} - \alpha$ (see Fig. \ref{fig:angles}). $\alpha$ denotes the angle between the interface normal (n) and the $k_\mathrm{x}$-direction and $\theta_\mathrm{S}$ is the angle between the interface normal and the momentum of the ELQ in the superconductor.

The zero-energy Andreev bound state (ZEBS) with $E = 0$ occurs if the phase difference of the pair potentials $\Delta_{+}$  and $\Delta_{-} $, denoted by $\Phi_+ - \Phi_-$ (where $\Delta_{\pm} = \abs{\Delta_{\pm}}  \exp (i \Phi_{\pm})$), in Eq. \eqref{eq:peak_condition} is $\pm \pi$. $\Phi_{\pm}$ simplifies to $\Phi_{\pm}=\phi_{\pm}$, if $k_{\mathrm{F}}$ does not depend on $\phi$ and a d+id pair potential with equal mixing of d$_{\mathrm{x}^2-\mathrm{y}^2}$ and d$_{\mathrm{xy}}$ on the square lattice is considered. The novel aspect of more harmonics and an angle dependent $k_{\mathrm{F}}$ leads to a more complicated peak structure in the differential conductance curves (see Figs. \ref{fig:graphene_short}, \ref{fig:cobaltates} and Supplementary Figure 1). One possibility to fulfill $\Phi_+ - \Phi_- = \pm\pi$ is $\Delta_+ = - \Delta_-$ for purely real pair potentials. The resulting ZEBS are seen in our fRG + STM calculations for the f-pairing phase in Fig. \ref{fig:f_short}, as well as in Fig. \ref{fig:f_cobaltates}.

However, for a complex d+id pairing potential, the condition $\Delta_+ = - \Delta_-$ is not sufficient. This is exactly the situation encountered for the inner sunflower structure of the d+id order parameter shown in Fig \ref{fig:did_short}: Since $\Delta_+ = \Delta_{\mathrm{x}^2-\mathrm{y}^2}(\theta_\mathrm{S} - \alpha) + i \Delta_{\mathrm{xy}}(\theta_\mathrm{S} - \alpha)$ and
$\Delta_- = \Delta_{\mathrm{x}^2-\mathrm{y}^2}(\pi-\theta_\mathrm{S} - \alpha)  + i \Delta_{\mathrm{xy}}(\pi-\theta_\mathrm{S} - \alpha)$, the phase difference between the two pair potentials is not restricted to multiples of $\pi$. Thus, the peak position moves between 0 and
$\text{min} (\abs{\Delta_+}, \abs{\Delta_-})$, depending on the relation of $\Delta_+$, $\Delta_-$ and the angle $\alpha$. This peak is also called double-split peak, since the zero energy conduction peak is split into two peaks positioned symmetrically at positive and negative finite energies. This confirms the usefulness of the analogy with the pseudo quantum well and implies that indeed the quasiparticles in the quantum well are only confined if their energies are less than the amplitudes of both pair potentials.

Our differential conductance for the full (fRG + STM) calculations of the d+id order parameter on the honeycomb lattice is very different 
from what is usually done in a phenomenological STM approach.
There, one considers only the first harmonic in the superconducting (SC) gap, in other words one assumes only short-range  (nearest-neighbour)  electron-electron interactions entering the pairing (gap) function. 
The "phenomenological", nearest-neighbour, dI/dV plot is shown in Fig.~2 of the Supplementary Information for the in-plane setup (see also Supplementary Discussion 2) and is even qualitatively different from the one obtained in Fig.~4a of this manuscript 
for the realistic  dI/dV signal, here for cobaltates. 
The difference between realistic material-oriented and phenomenological approaches comes from the fact that although the high-energy Hamiltonian (for the cobaltates, Eq. \ref{eq:H_cobaltates} of Sect. IV) includes only on-site electron-electron interactions,
the fRG flow will induce longer range electron-electron pairings beyond nearest-neighbours in the pairing channel.
Therefore, the full (fRG + STM) calculations  include additionally also these higher-order interactions (higher harmonics).
They give rise to a quite complex "sunflower" structure of the d+id order parameter. 
Therefore, the fRG procedure provides qualitatively new insight into the differential conductance and is obviously essential if one wants to uniquely resolve a topological superconductor (SC) state.

In contrast to the conductance curves of the d+id pairing phases, that reveal clear signatures of broken TRS, the conductance curves of the f pairing phases contain zero energy peaks, which are present due to conserved TRS. These zero energy peaks (ZEBS) are seen in the cuts (Fig. \ref{fig:f_short}). Additionally, the inset in Fig. \ref{fig:f_short} shows a zoom for small quasiparticle energies, displaying the ZEBS in white. As discussed already above, these peaks originate from an antisymmetric pairing $\Delta_+ = -\Delta_-$. 
The physical reason for this effect is a lack of inversion symmetry for the f order parameter with respect to the origin of the k$_\mathrm{x}$ - k$_\mathrm{y}$ plane (let us mention that this inversion is preserved for both real and imaginary parts of the d+id order parameter).
This lack of inversion symmetry and the corresponding rotation of the f order parameter can be easily understood, considering for example graphene (Fig. \ref{fig:f_short}) at an angle $\alpha=\frac{\pi}{6}$. Then, ELQ exhibit a pair potential of $\Delta_+=\Delta(\theta_\mathrm{S}- \frac{\pi}{6})$ and HLQ of $\Delta_{-} = \Delta(\frac{5\pi}{6}-\theta_\mathrm{S})$. Using the pair potential (the signs of the pair potentials for a phase difference of $\pi$ are opposite in Fig. \ref{fig:f_short}), we find $\Delta_+=-\Delta_-$, giving rise to a zero energy peak, which is indeed  found in the conductance spectrum at $\alpha=\frac{\pi}{6}$.

Let us add a few more details, concerning the results in Figs. \ref{fig:graphene_short} and \ref{fig:cobaltates}.
Figure \ref{fig:graphene_short} shows the differential conductance spectra and the pairing potentials for d+id and f pairing phases of graphene with short-range Coulomb interactions at the van Hove singularity where the screening is very effective. Corresponding dI/dV characteristics for a larger doping $x = 0.15$ (long-range Coulomb interactions) are presented in Supplementary Figure 1 (see also Discussion).

A given $\alpha$-direction corresponds to a specific surface, which can be expressed with Miller indices.  Cross sections for $\alpha = 0$ and $\alpha = \frac{\pi}{6}$ correspond to Miller indices $(1,-1,0)$ and $(1,0,0)$ for cobaltates and, respectively, $(1,-1)$ and $(1,0)$ for graphene.

The cross sections of  dI/dV characteristics in Figs. \ref{fig:graphene_short} and \ref{fig:cobaltates} show, that the width of the zero energy peaks is maximal in the maximum bandgap directions of the pairing potential (given by $\alpha=\frac{(2n+1)\pi}{6}$, $n \in \mathbb{Z}$), in which the condition of antisymmetric pairing is fulfilled for all incident angles $\theta$  (angle between the incident momentum and the interface normal, see Sect. IV). In general, the height of the peaks depends on how many incident angles $\theta$  contribute to the resonance for a given $\alpha$-direction. In the case of the cobaltates, the gap is very anisotropic and more harmonics contribute than for graphene. Consequently, the number of incident angles contributing to a resonance is smaller, giving a smaller peak height.

Summarizing, the in-plane setup is sensitive to the magnitude and the phase of the pair potential, and allows to distinguish the different pairing phases (d+id and f). While the f-wave paring phase gives zero energy peaks (ZEBS) typical for order parameters with conserved TRS, the signature of broken TRS can be clearly seen in the sunflower structure in the differential conductance spectrum of the d+id paring phase of our prototypical examples.

\subsection*{Differential conductance spectra for the out-of-plane setup}


Figure \ref{fig:out} shows the differential conductance obtained in the out-of-plane setup for the d+id and f pairing phases of graphene and the cobaltates. The out-of-plane setup shows only the density of states features since the information about the phase and the in-plane angle dependence of the pairing potential is integrated out. The density of states induces peaks at energies that match the maximum pairing potential, causing maximal Andreev reflection and a peak in the dI/dV characteristics. For the d+id pairing phase of graphene there is only one peak at the maximum pairing potential ($8 \Delta$). On the other hand, the differential conductance curve of the d+id pairing phase of the cobaltates displays a number of peaks, each one corresponding to a local maximum (or saddle point) of the pairing potential. This signature is again a manifestation of the strongly anisotropic d+id pairing phase of the cobaltates.

Since the f pairing potentials are nodeless in some specific directions, the dI/dV curves for the f phases grow with the excitation energy until the peak for the maximal pairing potential is reached.
For the d+id pairing phase, a kink is observed at the minimum pairing potential value, because Andreev reflection occurs for some directions in momentum space.

Furthermore, for the d+id pairing phases, there is no conductance observed below the minimum pairing potential value, because normal electron reflection is the only process in this regime. 
This region of zero conductance, in many cases, is a signature of a gapped phase, 
that allows to distinguish between gapless and gapped pairing potentials. In our role-model system cobaltates ($\mathrm{T}_{\mathrm{C}} = 5\mathrm{K}$ \cite{takada_2003}), 
however, as shown in Fig.~3a,b of the Supplementary Information (differential conductance curves as a function of temperature for the out-of plane STM configuration)
this is not the case for temperatures of the order of a few Kelvins (0.25 $\Delta  \leq k_{\mathrm{B}}T <  k_{\mathrm{B}} \mathrm{T}_{\mathrm{C}}$). This  originates from the fact that the out-of plane setup is only sensitive to the density of states,
and the temperature broadening is of the order of the small d+id SC gap. 
In contrast, in the in-plane STM setup (see Figs.~3c and 3d of the Supplementary Information and Supplementary Discussion 3), the Andreev bound states in the d+id order parameter are robust against temperature smearing. 
That is exactly the reason why one needs not only the out-of-plane but also the in-plane STM setups to identify topological superconductivity in realistic materials.

\section{Discussion}

Our results show that the combination of the fRG pairing input with the STM spectroscopy analysis allows for an unambiguous characterization of the SC state and its pairing symmetry, starting from a microscopic, in principle \textit{a priori} description of the interacting Hamiltonian.

      The relevance of this combination becomes especially clear when considering the possibility of unconventional SC on hexagonal lattices.
      In many layered compounds, which are candidates for electronically driven (high T$_{\mathrm{C}}$) SC, the atoms form a square lattice. For the intensively studied square-lattice material classes, such as the cuprates and pnictides, experimental evidences and theoretical descriptions have already provided a rich picture. A celebrated example is, of course, the d-wave symmetry of the SC state in the cuprates, which by its momentum profile of the SC gap points directly to the decisive role of electronic correlations for the pairing mechanism.
      The situation is rather different for unconventional SC in hexagonal systems. In only a few hexagonal materials, the origin of SC can so far be assigned unambiguously to electronic interactions, partly due to the often occurring lattice distortions, which make a phonon-driven scenario of SC more likely. (It can even be such that electronic correlations strengthen phonon-induced pairing \cite{karakonstantakis_2013}.)

However, there are also compounds, where strong correlations, in combination with a hexagonal lattice symmetry are very likely to induce unconventional SC \cite{lefebvre_2000}.
    Bechgaard salts are certainly candidates for organic unconventional superconductors \cite{podolsky_2004, nickel_2005}. As mentioned before, the pnictide compound $\mathrm{SrPtAs}$ has recently attracted substantial attention:
 it is a multi-layer compound, where  Pt and As atoms are arranged in honeycomb rings.
Preliminary evidence for a TRS broken SC phase stems from $\mu$-SR data \cite{biswas_2013,fischer_2013}. Another relevant material class on the triangular lattice are the water-intercalated sodium cobaltates \cite{takada_2003}, which are discussed in detail here as one example
 for the strength of the fRG + STM method. Another promising avenue towards hexagonal Fermi-Surface (FS) instability  may be related to the emerging possibility of creating hexagonal optical lattices with fermionic isotopes \cite{jo_2012,uehlinger_2013} of ultra-cold atomic gases,
 provided that the limit $T < \mathrm{T}_\mathrm{C}$, $T/\mathrm{T}_\mathrm{F} \ll 1$, where $\mathrm{T}_\mathrm{F}$ is the Fermi temperature, can eventually be reached.

The cobaltates, very much like our other example graphene, for which the FS instability study is transferred from the triangular lattice of the cobaltates to the honeycomb lattice, constitute typical examples, where the fRG provides us with an approach to obtain the unbiased phase diagrams of the FS instabilities in all parquet channels.

The combination with the STM then elevates the microscopic theory, the fRG, to a new level directly accessible in scanning tunneling microscopy experiments.

This brings us finally to the discussion how to distinguish experimentally between different pairing phases using scanning tunneling microscopy. While one expects a gap in the dI/dV characteristics for the d+id phase (zero differential conductance until the minimal pairing potential is reached), the differential conductance increases continuously with energy from zero for the gapless f-pairing phase. However, the resolution in a STM experiment is mainly limited by the temperature. For realistic temperatures of $3\mathrm{K}$ to $4\mathrm{K} < \mathrm{T}_\mathrm{C}$ (for cobaltates), the resolution is typically around  $0.25\mathrm{meV}$ to $0.3\mathrm{meV}$.
If the superconducting energy scale $\Delta$ is of the order of 1meV, as expected from a rough estimate within fRG, it might be already very difficult to distinguish between d+id and f pairing phases for cobaltates in the out-of-plane STM setup (see Figs. 3a,b in the Supplementary Information). 
However, for the small gap of the highly anisotropic d+id pairing potential, a careful analysis of, additionally, the in-plane STM setup should help resolving 
all ambiguities. Here for both the cobaltates and graphene, one can clearly see the zero-energy peak (ZEBS) for the TRS-preserving SC state, in contrast to the d+id pairing phase where the characteristic inner sunflower structure is a fingerprint of the chiral SC state.

\section{Methods}

\subsection*{Details of the fRG calculations}

The strength of the fRG technique is evidenced in both graphene and cobaltate examples: both display near-nested Fermi surfaces \cite{kiesel_2012,kiesel_2013}, where SC has to compete with SDW and charge-density wave (CDW) instabilities.
The emergent orders are then determined in an unbiased manner by the RG flow of the corresponding susceptibilities and of the related interaction channels to low energies at the instability level, which is $\sim \mathrm{k}_\mathrm{B}\mathrm{T}_\mathrm{C}$ in the SC channel \cite{platt_2014}.

The high-energy starting point is given by the Hamiltonian
\begin{equation}
H = H_0 + H_{\mathrm{int}}
\label{eq:H}
\end{equation}
which is accurately determined for the band-structure part $H_0$ (typically taken from a fit to an a-priori DFT calculation) and for the interaction $H_{\mathrm{int}}$ (taken for example from a cRPA determination of the various terms). Details of the parameter choices can be found in Ref. \onlinecite{platt_2014} .

For the cobaltates, the Hamiltonian includes three hybridized orbitals per site (d$_{\mathrm{xy}}$, d$_{\mathrm{yz}}$, d$_{\mathrm{zx}}$) and reads \cite{bourgeois_2009}

\begin{align}
\begin{split}
H_{\text{eff}} = & \sum_{\langle i,j \rangle, \alpha, \beta, \sigma}((t+t'\delta_{\alpha,\beta}+D\delta_{ij})c^{\dagger}_{i\alpha \sigma}c_{j \beta \sigma}+\mathrm{h.c.}) \\
&+ \mu \sum_{i, \alpha, \sigma} n_{i \alpha \sigma} + U_1 \sum_{i, \alpha} n_{i\alpha \uparrow} n_{i \alpha \downarrow} \\
&+\frac{1}{2}\sum_{i, \alpha \neq \beta} (U_2 \sum_{\sigma, \nu} n_{i \alpha \sigma} n_{i \beta \nu} + J_H \sum_{\sigma, \nu}c^{\dagger}_{i \alpha \sigma}c^{\dagger}_{i\beta \nu}c_{i \alpha \nu}c_{i \beta \sigma} \\
&+ J_\mathrm{P} c^{\dagger}_{i \alpha \uparrow} c^{\dagger}_{i \alpha \downarrow}c_{i \beta \uparrow}c_{i \beta \downarrow}) ,
\end{split}
\label{eq:H_cobaltates}
\end{align}

where $c^{\dagger}_{i \alpha \sigma}$ denotes the electron creation operator with spin $\sigma=\uparrow, \downarrow$ and orbital $\alpha$ at site i. The occupation number is defined as $n_{i \alpha \sigma}=c^{\dagger}_{i \alpha \sigma}c_{i \alpha \sigma}$. In addition, t represents the hopping mediated by O$_{pz}$ orbitals and t’ corresponds to a direct Co-Co-hopping, D is the crystal-field splitting, and $\mu$ the chemical potential. These parameters are set to $t = 0.1 \mathrm{eV}$, $t’ = -0.02 \mathrm{eV}$, and $D = 0.10 \mathrm{eV}$.
The parameters $U_1 = 0.37 \mathrm{eV}$ and $U_2 = 0.25 \mathrm{eV}$ are intraorbital and interorbital Coulomb interactions, respectively. The remaining interaction parameters are $J_\mathrm{H} = J_\mathrm{p} = 0.07 \mathrm{eV}$  for Hund’s rule coupling $J_\mathrm{H}$ and pair hopping $J_\mathrm{p}$.

In graphene, the tight-binding Hamiltonian $H_0$ is
\begin{equation}
H_0=\left[t_1\sum_{\langle i,j \rangle, \, \sigma} c^{\dagger}_{i,\sigma}c_{j,\sigma}+t_2\sum_{\langle\langle i,j\rangle\rangle, \, \sigma} c^{\dagger}_{i,\sigma}c_{j,\sigma} +t_3 \sum_{\langle\langle\langle i,j \rangle\rangle\rangle, \, \sigma}c^{\dagger}_{i,\sigma}c_{j,\sigma} + \mathrm{h.c.} \right] -\mu n ,
\label{eq:H0_graphene}
\end{equation}
where $n=\sum_{i,\sigma} n_{i,\sigma}=\sum_{i,\sigma}c^{\dagger}_{i,\sigma}c_{i,\sigma}$. $c^{\dagger}_{i,\sigma}$ is the creation operator for an electron with spin $\sigma$ at site $i$,
$\mu$ denotes the chemical potential, and $t_{1\cdots 3}$ is the hopping strength for nearest neighbour (1), second nearest neighbour (2) and third nearest neighbour (3) hopping.
Coulomb interaction is included by a long-range Hubbard-type Hamiltonian $H_{\mathrm{int}}$ with
\begin{equation}
H_{\mathrm{int}}=U_0\sum_i n_{i,\uparrow}n_{i,\downarrow} +\frac{1}{2}U_1\sum_{\langle i,j \rangle, \, \sigma, \, \sigma'}n_{i,\sigma}n_{j,\sigma'}+\frac{1}{2}U_2\sum_{\langle\langle i,j \rangle\rangle, \, \sigma, \, \sigma'}n_{i,\sigma}n_{j,\sigma'} ,
\label{eq:Hint_graphene}
\end{equation}
where $U_{0 \cdots 2}$ gives the Coulomb repulsion scale from on-site (0) to second nearest neighbour (2) interactions, respectively.

\begin{figure}[p]
\centering
\includegraphics[width=\linewidth]{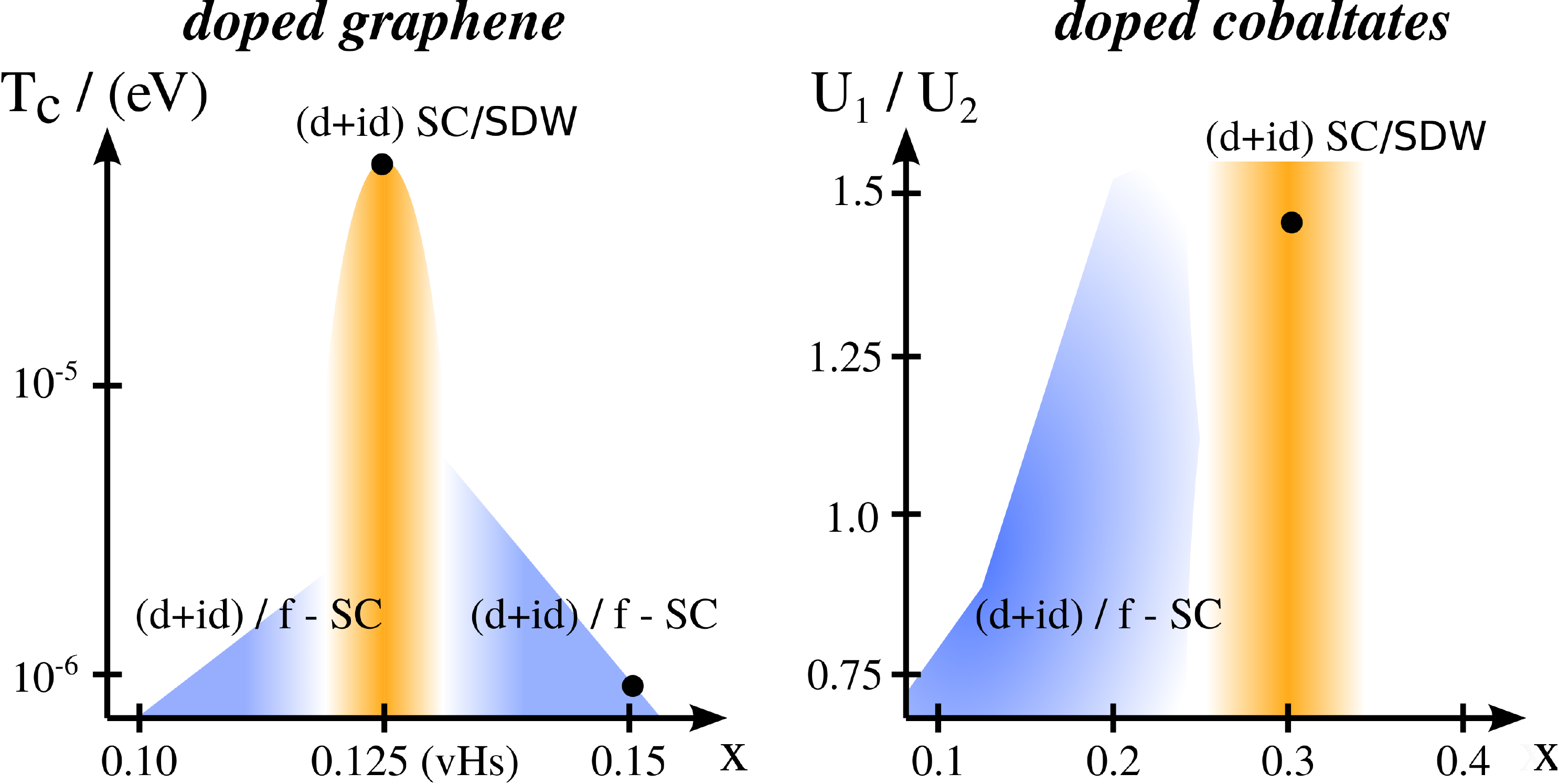}
\caption{Schematic phase diagrams obtained from fRG calculations (see Sect. IV) for doped graphene and water-intercalated sodium cobaltates. The competing phases  consist of the chiral superconducting d+id phase in the singlet channel, the f-wave phase in the triplet channel as well as the competition with a SDW phase. The black dots denote the phase-diagram points for which the TRS-breaking d+id-phase and the TRS-conserving f-phase are calculated and taken as inputs in the STM scheme. For the cobaltates, the doping has been confined for both d+id and f-phases to the value $x \cong 0.3$, where SC has been observed experimentally.}
\label{fig:phasescheme}
\end{figure}

\begin{figure}[p]
\begin{overpic}[width = \linewidth]{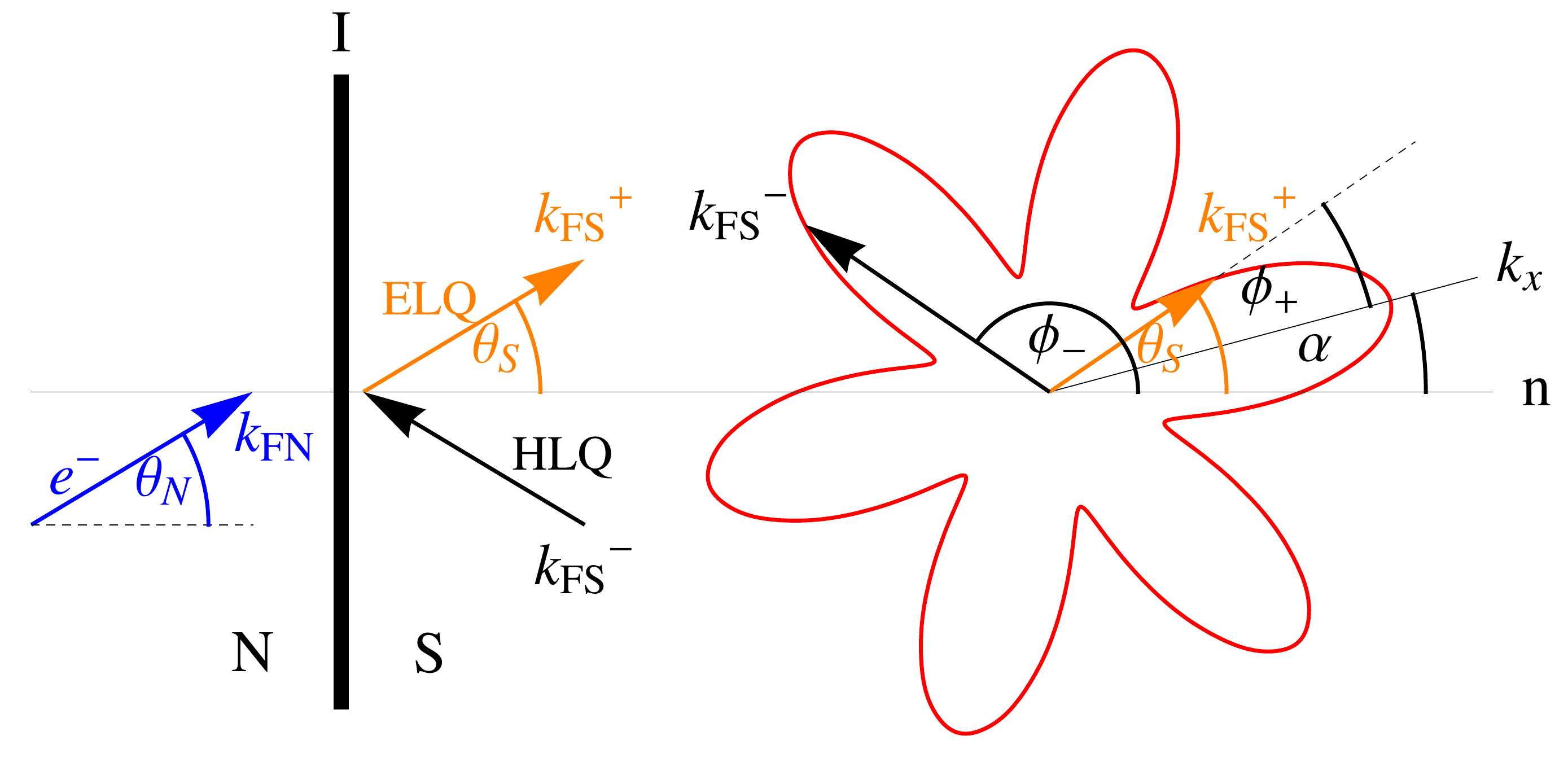}
   \put(0,30){\boxed{\includegraphics[width=0.19 \linewidth]{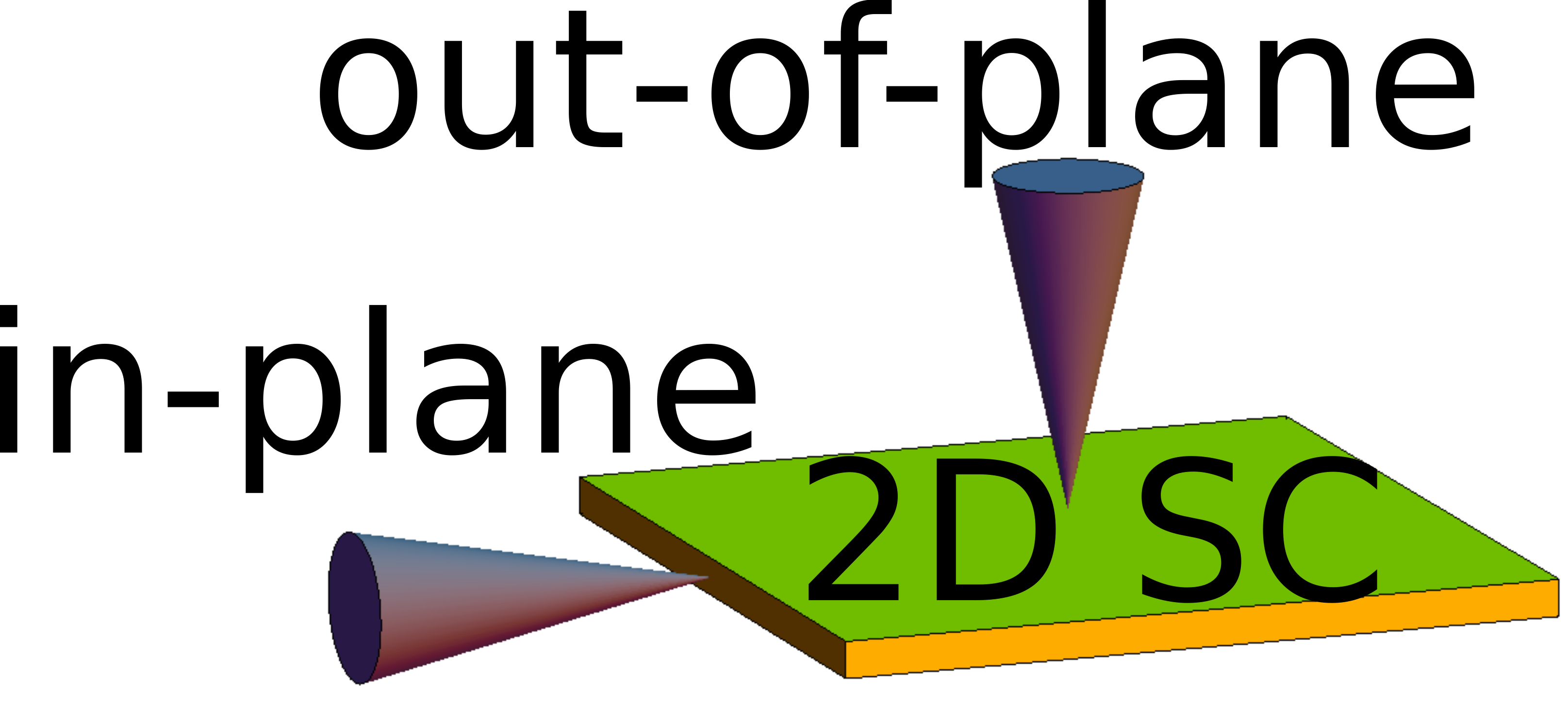}}}
\end{overpic}
\caption{Angle definitions for the normal metal (N) - insulator (I) - superconductor (S) junction. An incident electron on the normal side is characterised by the angle $\theta_{\mathrm{N}}$ (with respect to the surface normal n). On the superconducting side, transmitted electron-like quasiparticles (ELQ) have the momentum vector $k_{\mathrm{FS}}^+$ (angle $\theta_{\mathrm{S}}$ with respect to the surface normal n) and hole-like quasiparticles (HLQ) have $k_{\mathrm{FS}}^-$,  and, thus, experience  the corresponding different effective pair potentials. The $k_{\mathrm{x}}$-direction of the pair potential can be tilted by an angle $\alpha$ with respect to the surface normal n. The inset (box) shows the setup: The STM tip can be placed either in-plane or out-of-plane with respect to the quasi two-dimensional (2D) superconductor.
}
\label{fig:angles}
\end{figure}

\begin{figure}[p]
\subfloat{
\begin{minipage}[c]{0.59\linewidth}
\centering
\includegraphics[width=\linewidth]{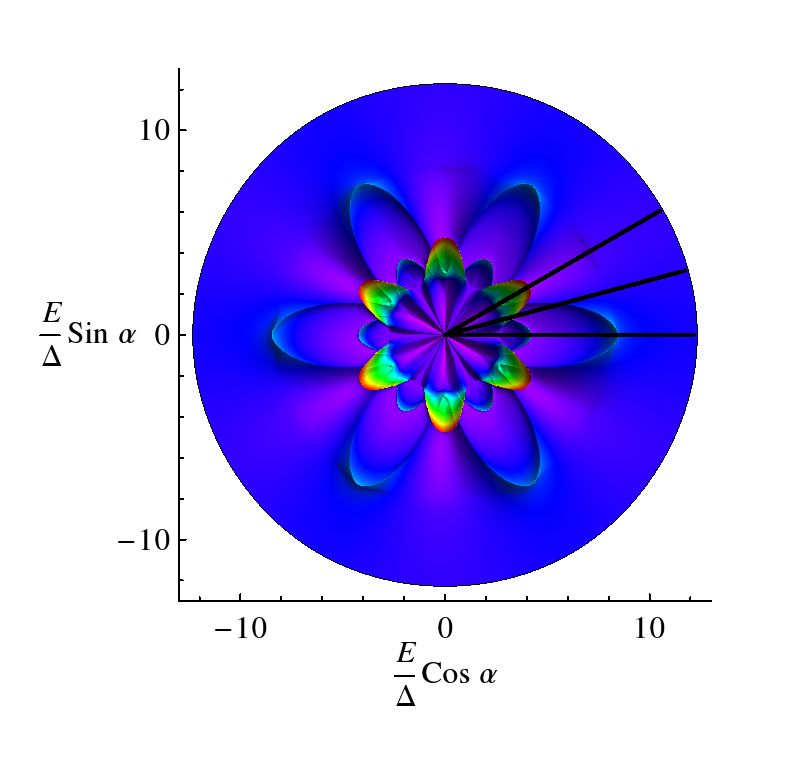}
\end{minipage}
\begin{minipage}[c]{0.05\linewidth}
 \Large \textbf{(a)} \vspace*{0.5cm}\\
\includegraphics[width=\linewidth]{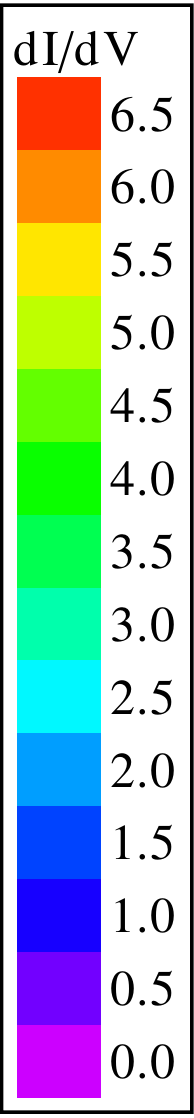}
\vspace*{1.5cm}
\end{minipage}
\begin{minipage}[c]{0.34\linewidth}
\hfill
\centering
\includegraphics[width=\linewidth]{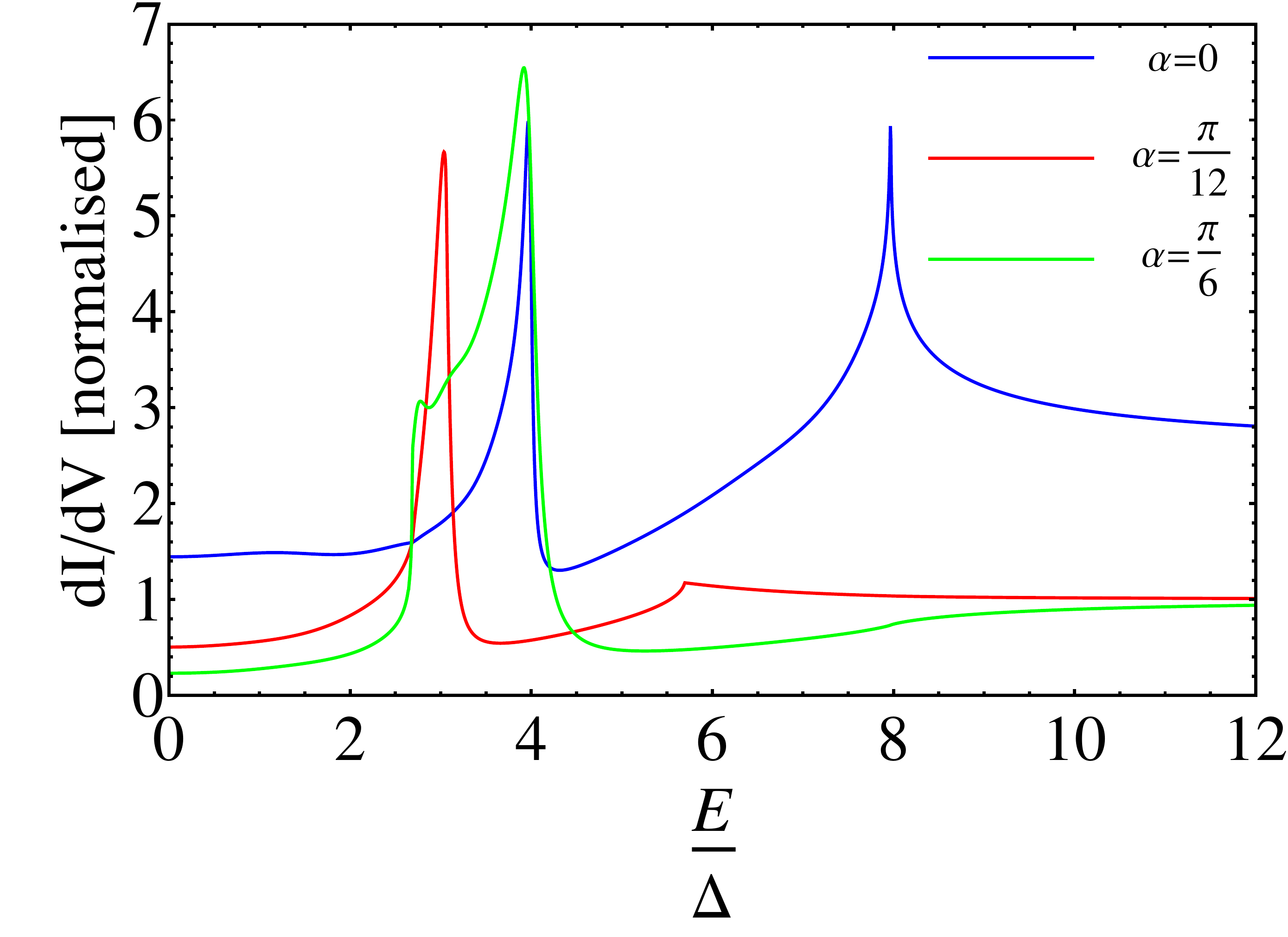}\\
\begin{minipage}[c]{0.53\linewidth}
\includegraphics[width=\linewidth]{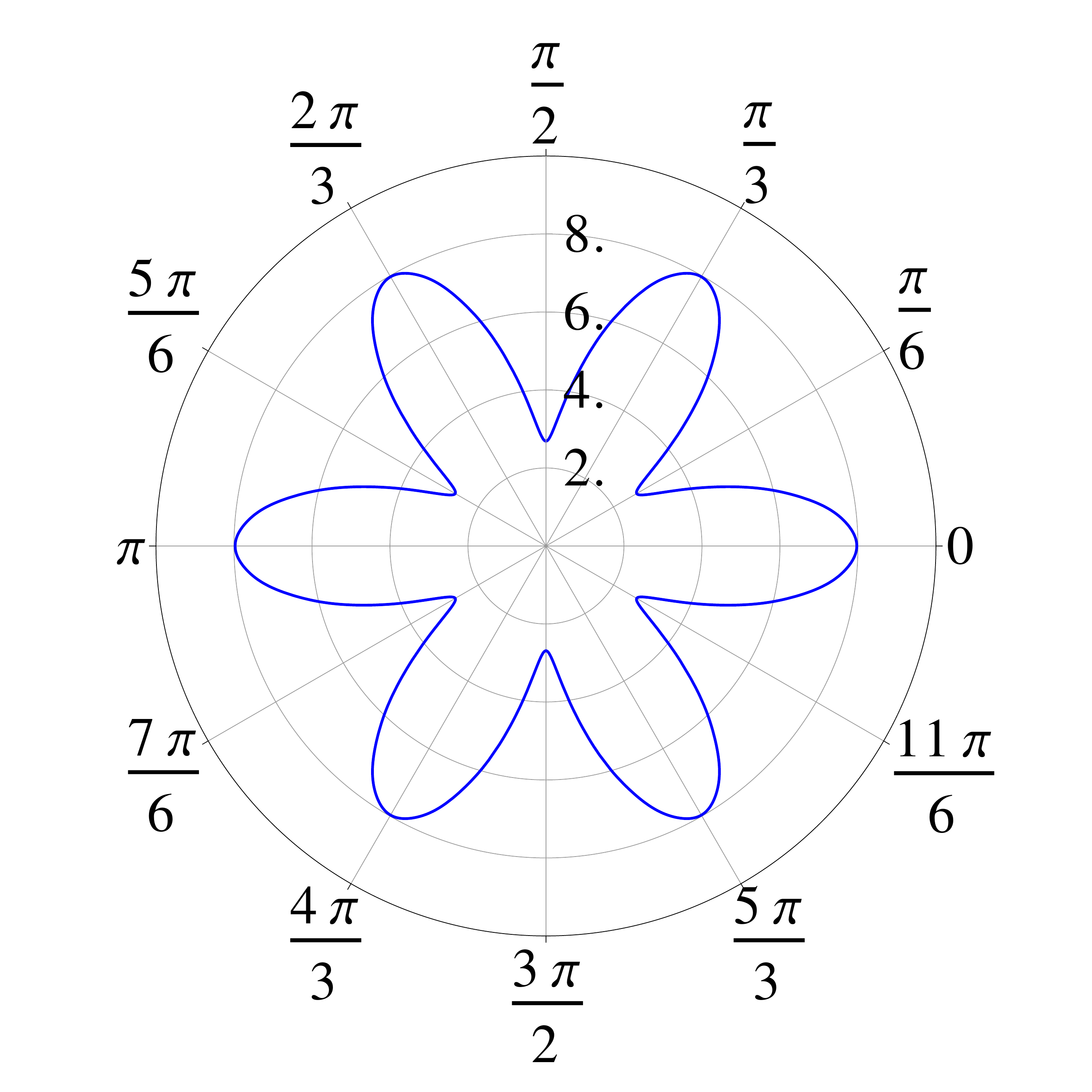}
\end{minipage}
\begin{minipage}[c]{0.45\linewidth}
\hfill
\end{minipage}
\end{minipage}
\label{fig:did_short}
}\\
\subfloat{
\begin{minipage}[c]{0.59\linewidth}
\centering
\includegraphics[width=\linewidth]{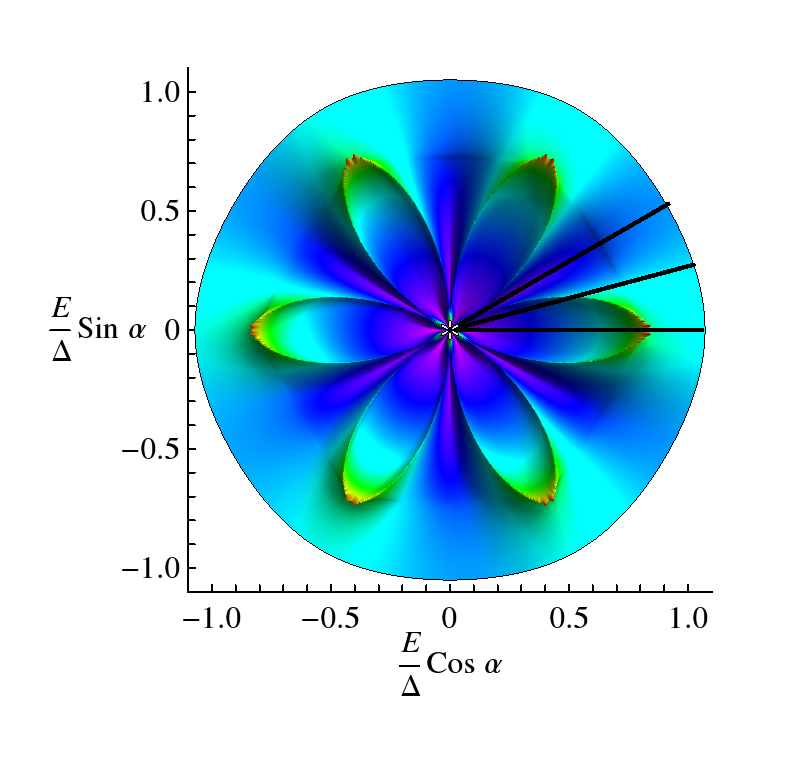}
\end{minipage}
\begin{minipage}[c]{0.05\linewidth}
 \Large \textbf{(b)} \vspace*{0.5cm}\\
\includegraphics[width=\linewidth]{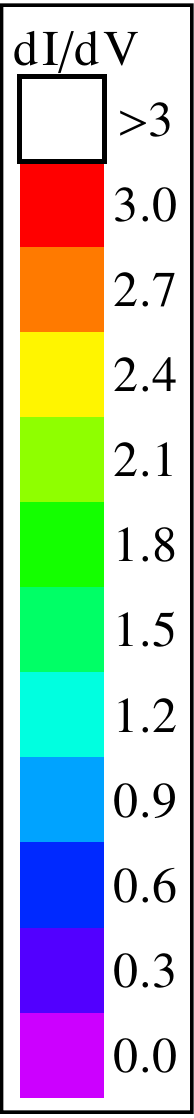}
\vspace*{1.5cm}
\end{minipage}
\begin{minipage}[c]{0.34\linewidth}
\centering
\includegraphics[width=\linewidth]{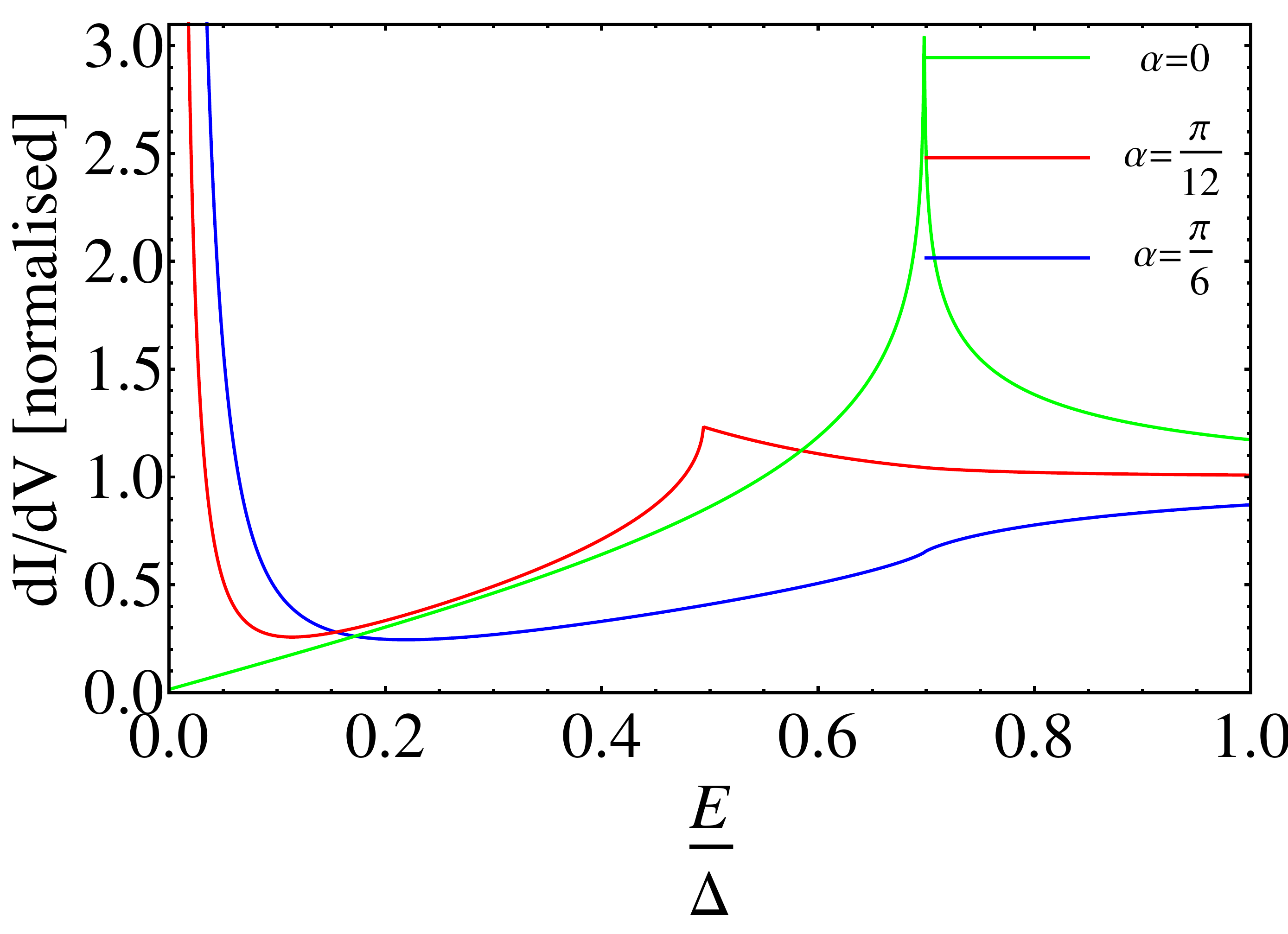}\\
\begin{minipage}[c]{0.53\linewidth}
\includegraphics[width=\linewidth]{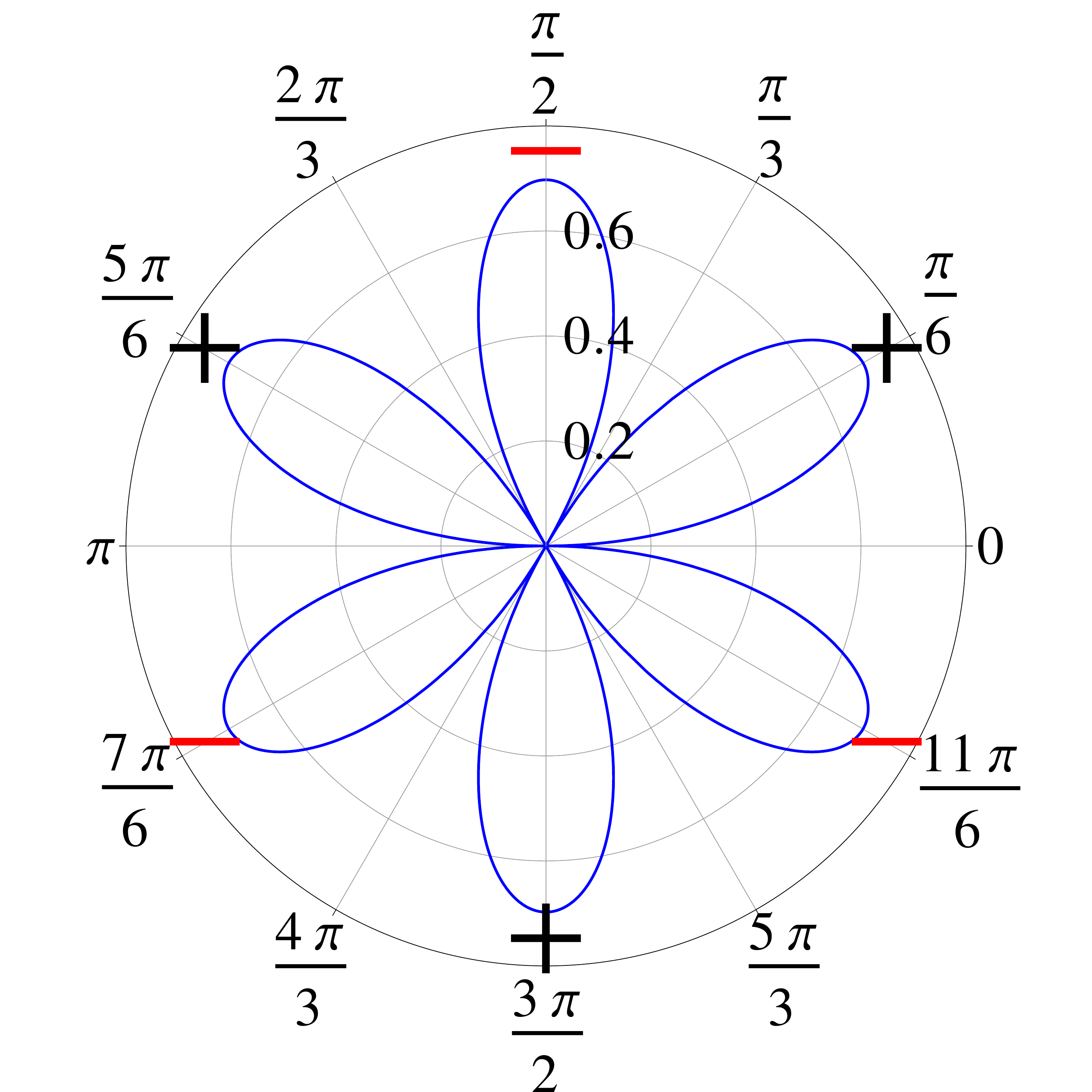}
\end{minipage}
\begin{minipage}[c]{0.45\linewidth}
\includegraphics[width=\linewidth]{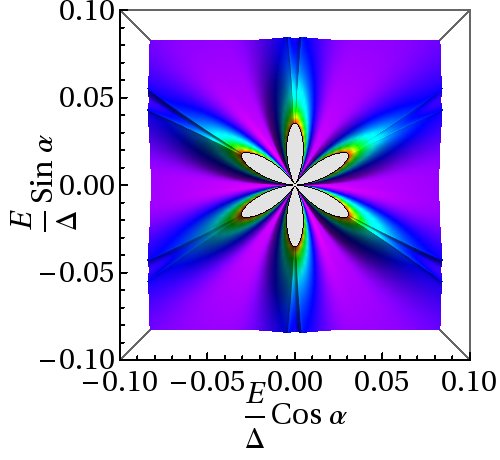}
\end{minipage}
\end{minipage}
\label{fig:f_short}
}
\caption{
Differential conductance spectra (dI/dV) and corresponding pairing potentials for graphene. The left panels show the differential conductance spectra for the d+id pairing phase (a) and the f pairing phase (b). The corresponding pairing potentials are shown on the right in the lower panels as a polar plot of the absolute value of the pairing potential. The differential conductance spectra, obtained in the in-plane setup simulating an STM experiment, are plotted as a function of the quasiparticle energy E (radial axis, normalised by the reference band gap $\Delta$) and the angle $\alpha$ between the interface normal and the $k_{\mathrm{x}}$-direction of the pairing potential (polar axis). Differences in brightness of the colors are due to the plot style. The black lines indicate angles $\alpha$, for which the cross sections are shown in the upper panels on the right. In (b) a zoom for small energies is presented, that shows the zero energy peaks.
}
\label{fig:graphene_short}
\end{figure}

\begin{figure}[p]
\subfloat{
\begin{minipage}[c]{0.59\linewidth}
\centering
\includegraphics[width=\linewidth]{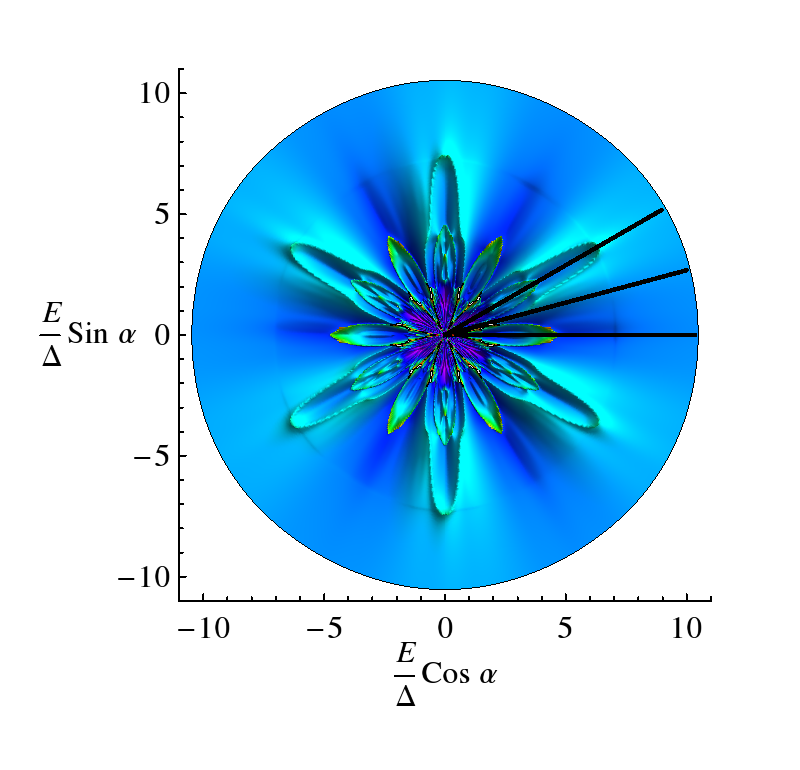}
\end{minipage}
\begin{minipage}[c]{0.05\linewidth}
 \Large \textbf{(a)} \vspace*{0.5cm}\\
\includegraphics[width=\linewidth]{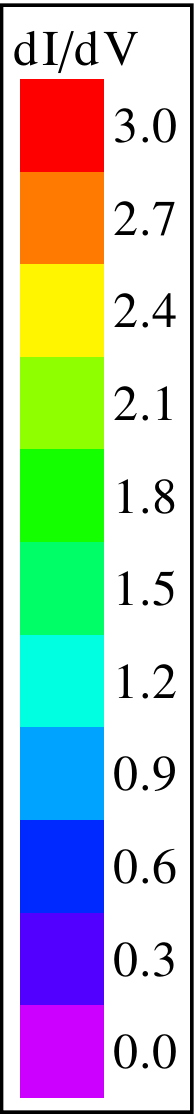}
\vspace*{1.5cm}
\end{minipage}
\begin{minipage}[c]{0.34\linewidth}
\centering
\includegraphics[width=\linewidth]{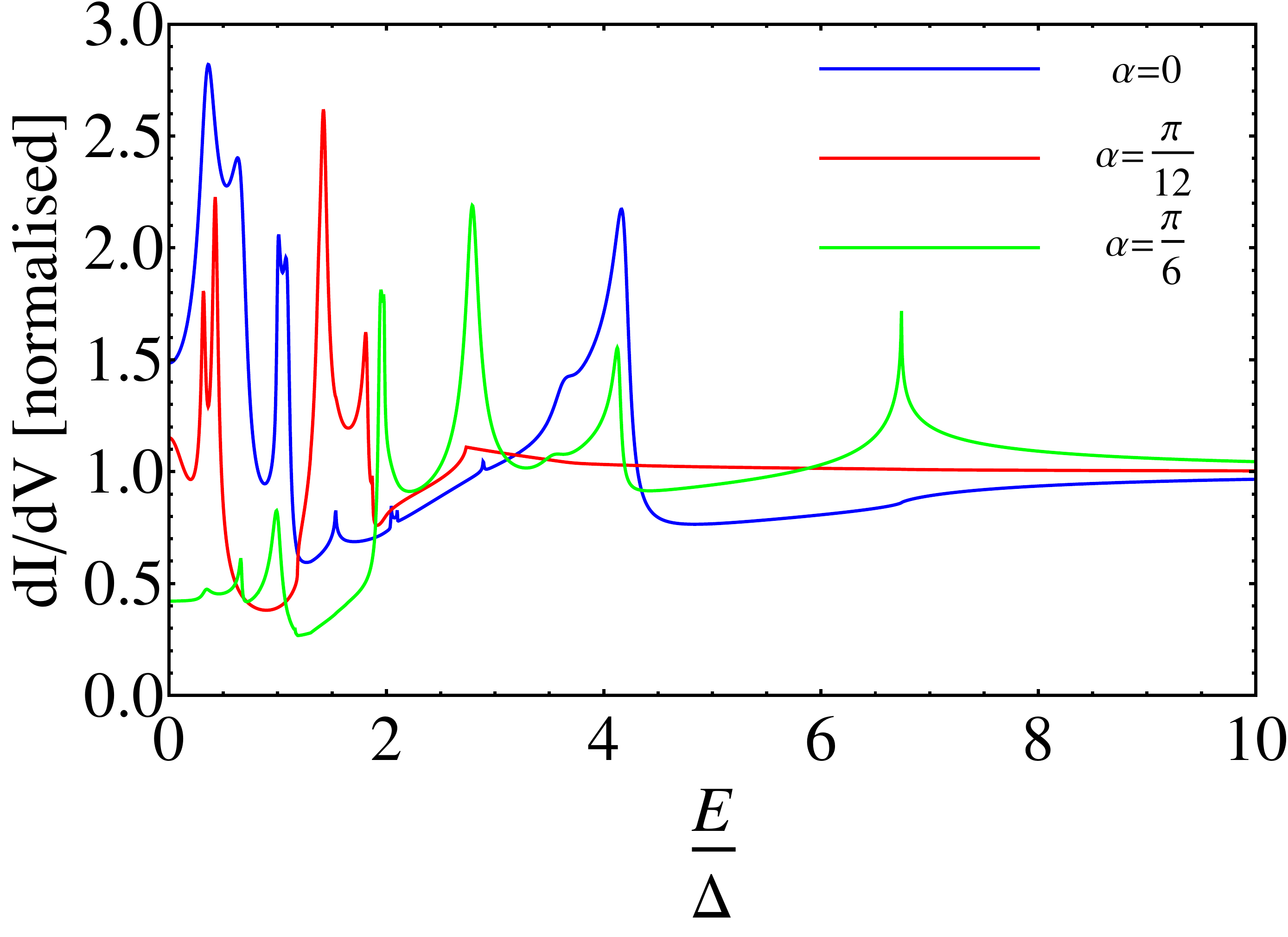}
\includegraphics[width=\linewidth]{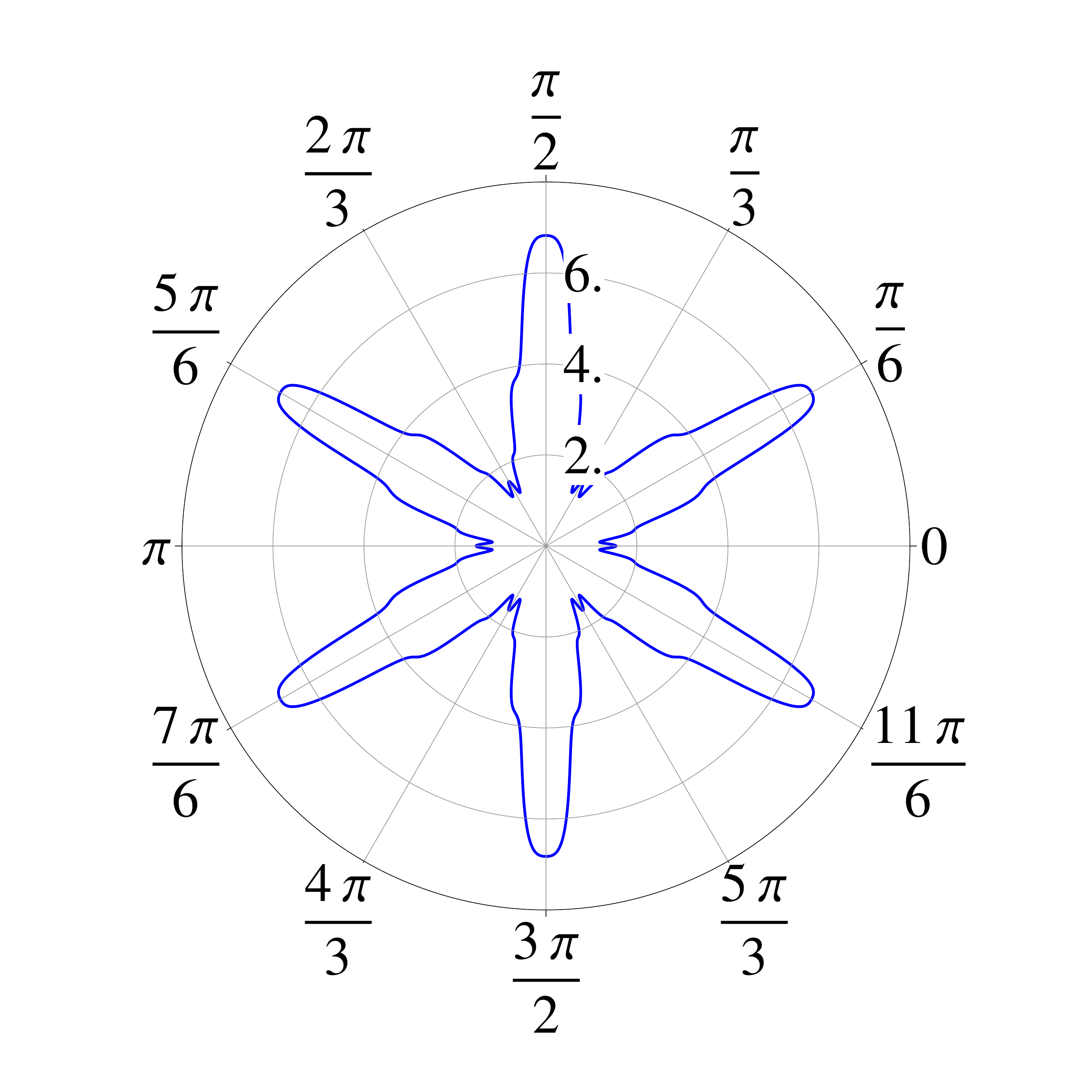}
\end{minipage}
\label{fig:did_cobaltates}
}\\
\subfloat{
\begin{minipage}[c]{0.59\linewidth}
\centering
\includegraphics[width=\linewidth]{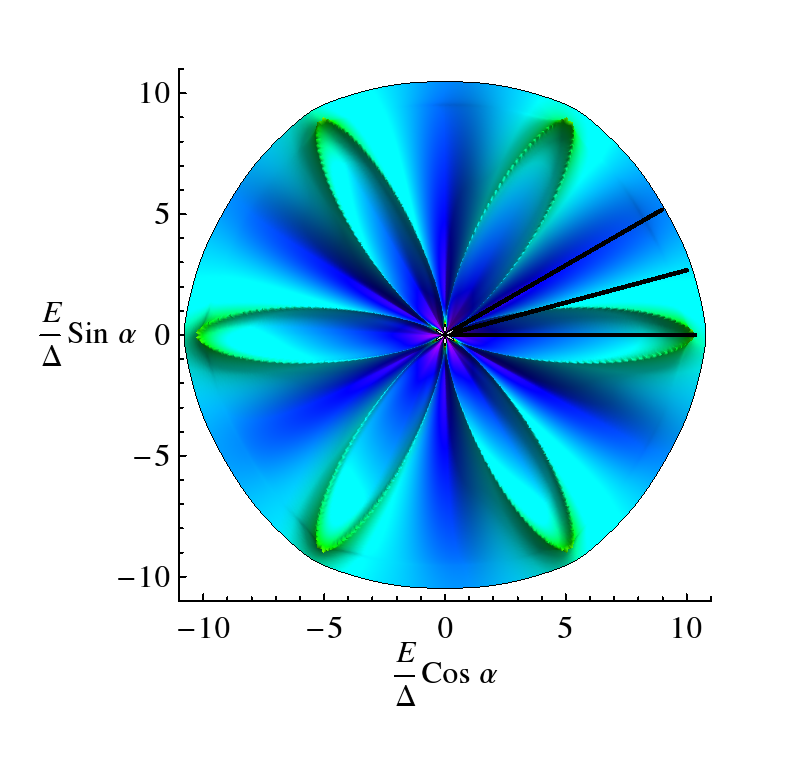}
\end{minipage}
\begin{minipage}[c]{0.05\linewidth}
 \Large \textbf{(b)} \vspace*{0.5cm}\\
\includegraphics[width=\linewidth]{Legend3_final}
\vspace*{1.5cm}
\end{minipage}
\begin{minipage}[c]{0.34\linewidth}
\centering
\includegraphics[width=\linewidth]{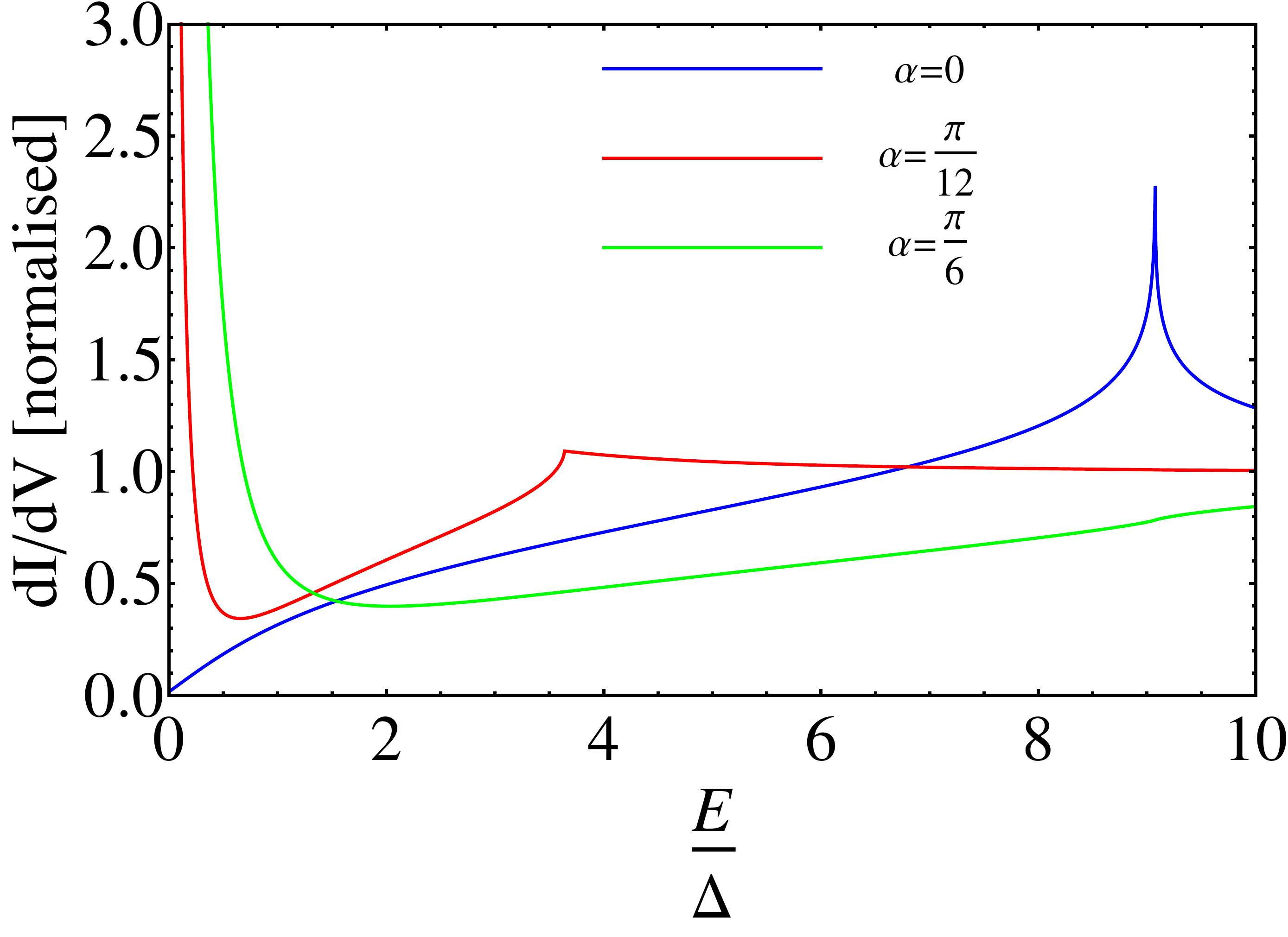}
\includegraphics[width=\linewidth]{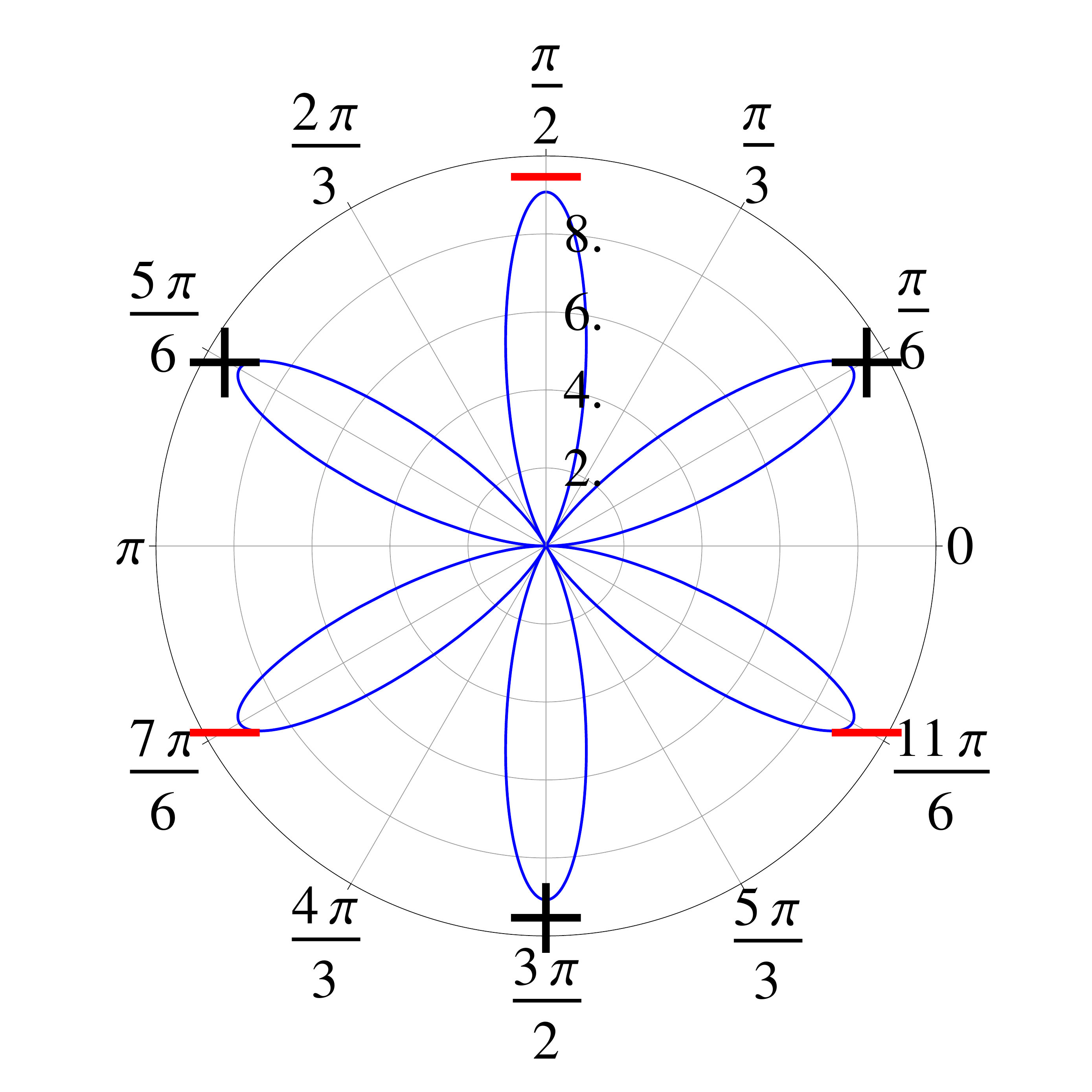}
\end{minipage}
\label{fig:f_cobaltates}
}
\caption{
Differential conductance spectra (dI/dV) and corresponding pairing potentials for the cobaltates. The left panels show the differential conductance spectra for the d+id pairing phase (a) and the f pairing phase (b). The corresponding pairing potentials are shown on the right in the lower panels as a polar plot of the absolute value of the pairing potential. The differential conductance spectra, obtained in the in-plane setup simulating an STM experiment, are plotted as a function of the quasiparticle energy E (radial axis, normalised by the reference band gap $\Delta$) and the angle $\alpha$ between the interface normal and the $k_{\mathrm{x}}$-direction of the pairing potential (polar axis). Differences in brightness of the colors are due to the plot style. The black lines indicate angles $\alpha$, for which the cross sections are shown in the upper panels on the right.
}
\label{fig:cobaltates}
\end{figure}

\begin{figure}[h]
\includegraphics[width=\linewidth]{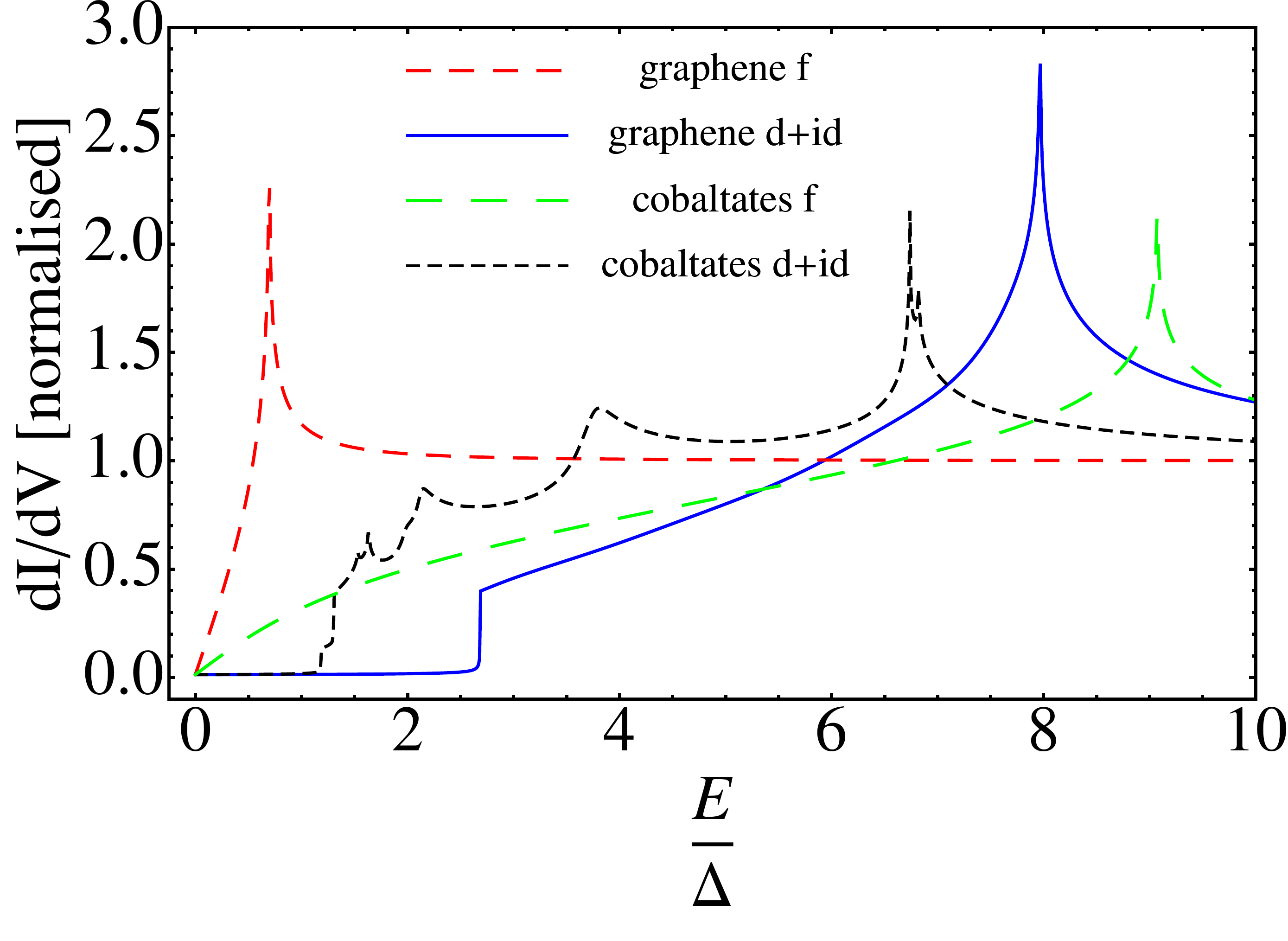}
\caption{Differential conductance (dI/dV) as a function of the quasiparticle energy E normalised by the reference bandgap $\Delta$ for the out-of-plane setup at $T=0$. The curves of the d+id and f pairing phases are given for graphene and the cobaltates.
}
\label{fig:out}
\end{figure}

The near-degeneracy between SC and density-wave orders is strongly influenced by a subtle interplay between deviations from perfect nesting (taken into account in $H_0$ of Eqs. \eqref{eq:H_cobaltates} and \eqref{eq:H0_graphene} via longer-ranged hopping terms). Similarly, the near-degeneracy between TRS-breaking d+id SC order and TRS-preserving f-wave SC order is affected both by the Fermi surface topology and by the interaction terms. For example, in graphene at the vHs, we assume perfect screening and consider a Hubbard-type of Hamiltonian with on-site interaction $U_0 = 10 \mathrm{eV}$. Here a d+id SC phase is found (Fig. \ref{fig:phasescheme}). Away from the vHs ($x = 0.125$), we take longer-ranged Coulomb interactions into account [$U_1/U_0 = 0.45$, $U_2/U_0 = 0.15$]. The latter interactions determine, in particular, whether the competing f-wave SC instability is preferred.
      Using the above Hamiltonian (Eq. \eqref{eq:H}), we then employ the fRG and study how the renormalised interaction evolves under integrating out high-energy fermionic modes. At weak to moderate electron-electron interactions, this flow is accurately described by the fRG method, where one considers the flow of a function f (instead of a parameter) such as the interaction vertex, depending on 4 momenta \cite{platt_2014}. The renormalised interaction vertex (the 4-point function) is $V^{\Lambda}(k_1, k_2, k_3, k_4)$, where the flow parameter $\Lambda$ corresponds to the effective or low-energy scale temperature and $k_i$ label the incoming and outgoing momenta and the associated band indices.
     The starting conditions of the RG are given by the bare interactions as contained in Eqs. \eqref{eq:H_cobaltates} and \eqref{eq:Hint_graphene}, at an energy scale of the order of the bandwidth. Following the flow (we are using the temperature-flow fRG \cite{Honerkamp_2001})
 of the 4-point function (4 PF) $V^{\Lambda}$ down to low energies, the diverging channels at $\Lambda_\mathrm{C}$ then signal the nature of the instability, with $\Lambda_\mathrm{C}$ providing an upper bound for T$_\mathrm{C}$. 
At this low-energy scale, the flow has to be stopped and the remaining modes be treated with a different approach.
We resort to a mean-field scheme, where the effective interaction determines the SC gap function. This is a standard procedure, which has been used in many applications (see, for example, Ref. \onlinecite{platt_2014}). 

The phase diagram for graphene is plotted in Fig. \ref{fig:phasescheme}a. It displays the critical instability scale $\Lambda_\mathrm{C} \sim \mathrm{T}_\mathrm{C}$ as a function of doping. The phase diagram for the water-intercalated sodium cobaltates is presented in Fig.~\ref{fig:phasescheme}b 
as a function of doping and interaction ratio. We note that, when the nesting of the FS is optimal, strong SDW fluctuations along with singlet d+id SC occur. On the other hand, in the proximity of ferromagnetic fluctuations (FM) (which appear in the cobaltate case in Fig. \ref{fig:phasescheme} for large interaction 
ratios $U_1/U_2$, where the large DOS at the vHs promotes fluctuations with zero-momentum transfer),  triplet SC with a f-wave gap form factor (fSC) is dominant.

The 4PF $V_{\Lambda}(k,-k,q,-q)$ in the Cooper channel is, then, decomposed into different eigenmode contributions \cite{platt_2014}.
\begin{equation}
W^{\Lambda, \mathrm{SC}}(k,p) = \sum_i w_i^{\mathrm{SC}}(\Lambda)f_i^{\mathrm{SC}}(k)^*f_i^{\mathrm{SC}}(p) ,
\end{equation}
where i is a symmetry decomposition index. The leading instability of that channel corresponds to an eigenvalue $w_i^{\mathrm{SC}}(\Lambda)$ first diverging under the flow of $\Lambda$. $f_i^{\mathrm{SC}}(k)$ is the SC form factor of pairing mode i, which tells us about the SC pairing symmetry and hence gap structure associated with it.  In the fRG, from the final Cooper channel 4PFs, this quantity is computed along the discretized FSs. $f_i^{\mathrm{SC}}(k)$ is the gap function which enters the BTK formula \eqref{eq:btk} for the conductance (as an example the harmonics for graphene with the short-range electron-electron interactions are shown in Supplementary Methods 1).

\subsection*{Details of the STM calculations}
For the STM-part of the calculation, we use the NIS junction setup and solve the Bogoliubov-de Gennes  equations for the normal and the superconducting parts. The insulator is modelled by a delta-Dirac potential barrier with strength H \cite{kashiwaya_1996}. We apply the approximation, that the quasiparticle excitation energy E and the maximum absolute value of the pairing potential $\mathrm{max}\{\abs{\Delta}\}$ are much smaller than the Fermi energy E$_\mathrm{F}$. Consequently, the pairing is only relevant close to the Fermi surface. Within these approximations, the wave function obtained by solving the BdG equations (see Supplementary Methods 2) is independent of the dispersion relation of the material (we assume to have a quadratic term in the dispersion).
 We use the continuity of the wave function at the interface and the matching of the derivative of the wave functions of the normal and the superconducting side to obtain the coefficients for Andreev ($r_\mathrm{A}$) and normal ($r_\mathrm{N}$) reflection.
The total conductance of the NIS junction is obtained by integrating the BTK conductance over all independent contributions, which is integrating over all k$_\mathrm{y}$ in our in-plane setup. We normalise the conductance by the conductance of a NIN junction in the same geometrical setup.

A peak in the conductance spectrum is obtained, if the following condition, first reported by Kashiwaya \textit{et. al.} \cite{kashiwaya_1996} is fulfilled
\begin{equation}
\Gamma_+ \Gamma_- = e^{i (\Phi_+ - \Phi_-)} ,
\label{eq:peak_condition}
\end{equation}
where $\Gamma_{\pm} = \frac{E-\sqrt{E^2-\absq{\Delta_{\pm}}}}{\abs{\Delta_{\pm}}}$, $\Phi_{\pm}= \mathrm{arg} (\Delta_{\pm})$ and E denotes the quasiparticle excitation energy. In general, $\Phi_{\pm}\neq \phi_{\pm}$ because k$_\mathrm{F}$ depends on $\phi$. However, for the generic d+id pair potential, $\Phi_{\pm}= \phi_{\pm}$.

\bibliography{literature/lit_final_9}

\section*{Acknowledgement}
We acknowledge financial support by the German research foundation (DFG) within FOR1162 (HA5893/5-2 and HA1537/23-2), within  FOR1458/2,  within SPP (HA1537/24-2), within SPP 1666 (HA 5893/4-1), 
and by the European Research Council within ERC-StG-TOPOLECTRICS-336012. We thank Matthias Bode and Ali Yazdani for useful discussions.

\clearpage

{\centering \large \textbf{Supplementary Information: Accessing  topological superconductivity via a combined STM and renormalization group analysis}}

\setcounter{equation}{0}
\setcounter{figure}{0}
\makeatletter
\renewcommand{\theequation}{S\arabic{equation}}
\renewcommand{\figurename}{Supplementary Figure}

\section*{Supplementary Discussion 1: differential conductance for graphene with long-range Coulomb interactions}
\begin{figure}[p]
\subfloat{
\begin{minipage}[c]{0.59\linewidth}
\centering
\includegraphics[width=\linewidth]{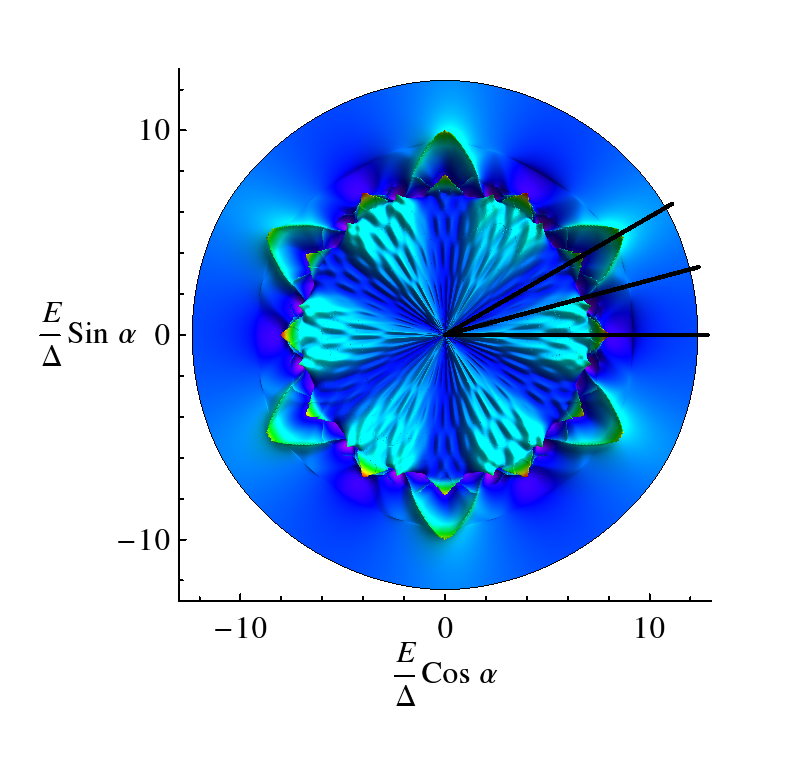}
\end{minipage}
\begin{minipage}[c]{0.05\linewidth}
 \Large \textbf{(a)} \vspace*{0.5cm}\\
\includegraphics[width=\linewidth]{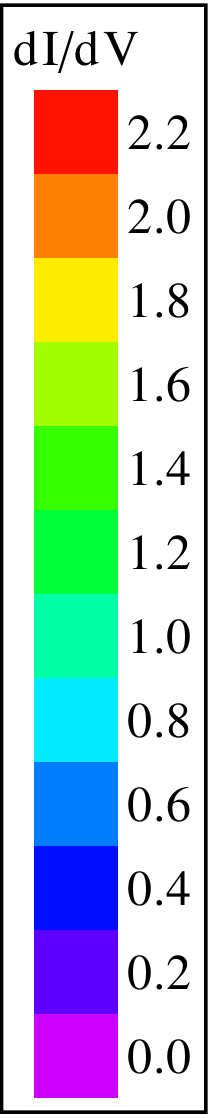}
\vspace*{1.5cm}
\end{minipage}
\begin{minipage}[c]{0.34\linewidth}
\centering
\includegraphics[width=\linewidth]{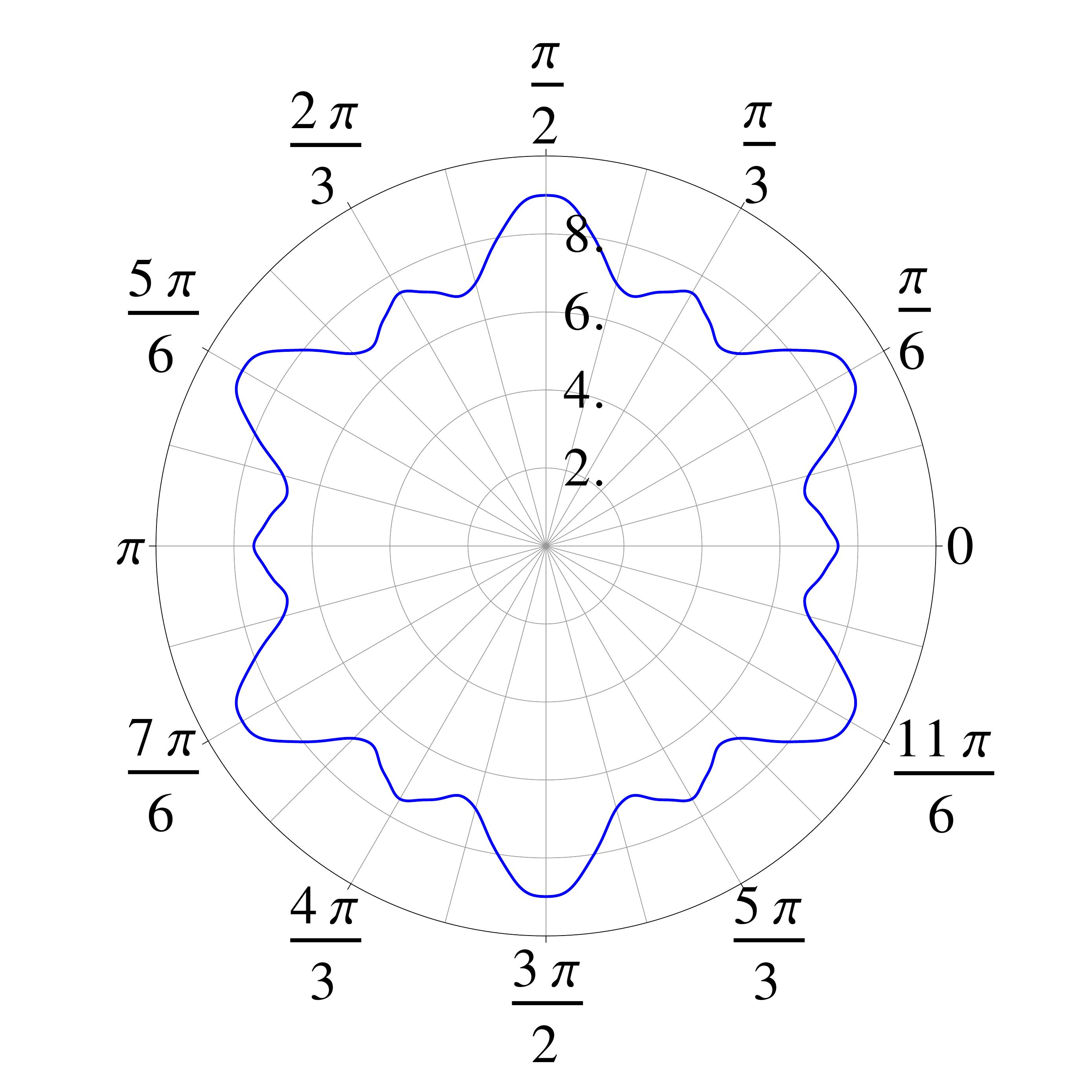}
\end{minipage}
\label{fig:did_long}
}\\
\subfloat{
\begin{minipage}[c]{0.59\linewidth}
\centering
\includegraphics[width=\linewidth]{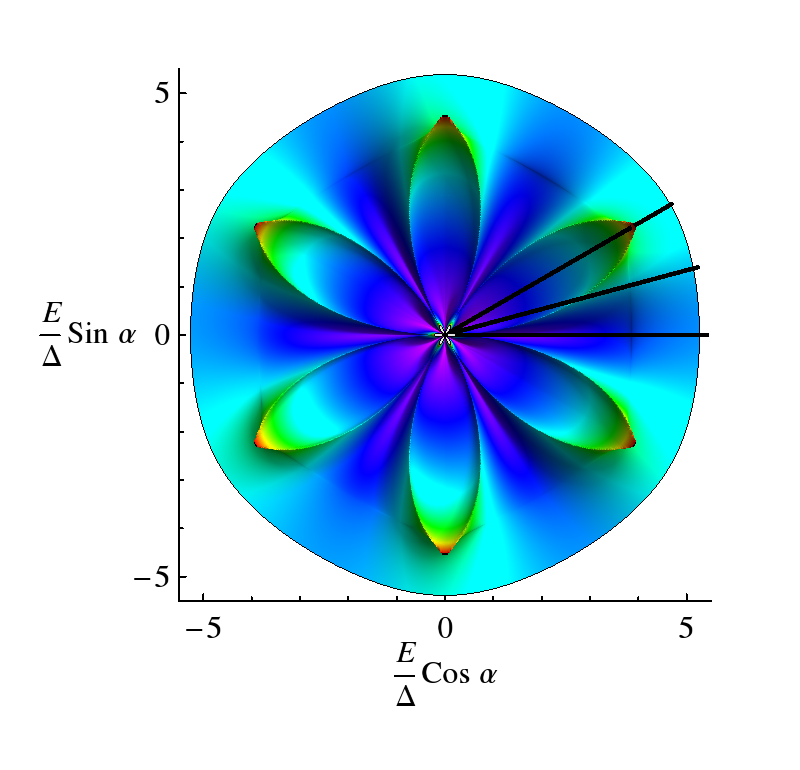}
\end{minipage}
\begin{minipage}[c]{0.05\linewidth}
 \Large \textbf{(b)} \vspace*{0.5cm}\\
\includegraphics[width=\linewidth]{Legend3_final}
\vspace*{1.5cm}
\end{minipage}
\begin{minipage}[c]{0.34\linewidth}
\centering
\includegraphics[width=0.8\linewidth]{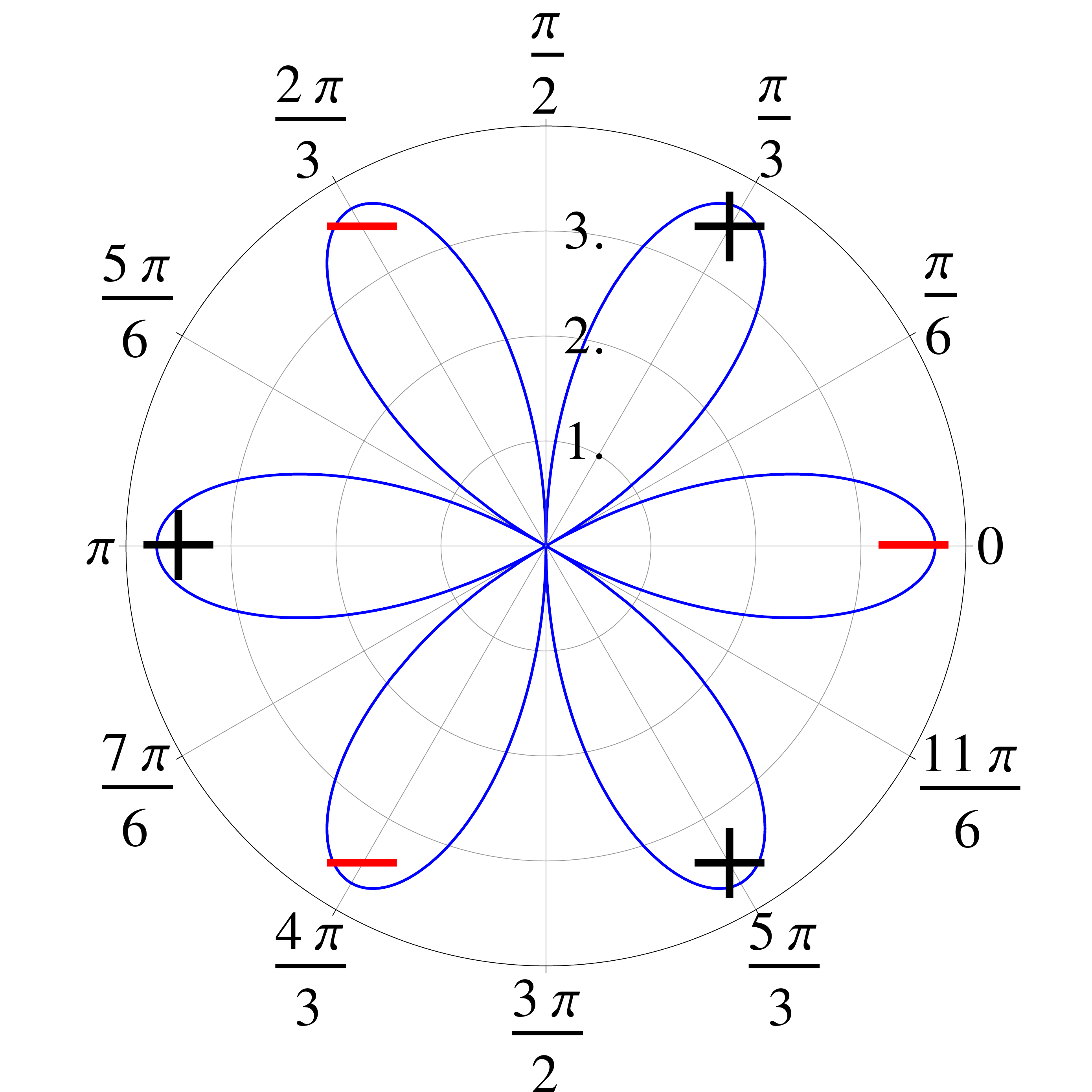}
\end{minipage}
\label{fig:f_long}
}
\caption{
Differential conductance spectra (dI/dV) for the in-plane setup and corresponding pairing potentials for the long-range Coulomb interaction pairing phase of graphene. The left panels show the differential conductance spectra for the d+id pairing phase (a) and the f pairing phase (b). The corresponding pairing potentials are shown on the right as a polar plot of the absolute value of the pairing potential. The differential conductance spectra, obtained in the in-plane setup simulating an STM experiment, are plotted as a function of the quasiparticle energy E (radial axis, normalised by the reference band gap $\Delta$) and the angle $\alpha$ between the interface normal and the $k_{\mathrm{x}}$-direction of the pairing potential (polar axis). Differences in brightness of the colors are due to the plot style. 
}
\label{fig:graphene_long}
\end{figure}
Fig.~3 in the main part of the manuscript displays the in-plane setup differential conductance spectra for the short-range Coulomb interactions in graphene at the van-Hove singularity.

Supplementary Fig. \ref{fig:graphene_long}a depicts the corresponding spectra and pairing potentials for graphene with longer-range Coulomb interactions for the d+id phase ($U_0 = 10\mathrm{eV}$, $U_1 =4.5\mathrm{eV}$ and $U_2 = 1.5\mathrm{eV}$, see Eq.~(5) in Sect. IV of the main part), which naturally appears, because away from the VHS (here $x = 0.15$) the screening is less effective.
As in  Figs.~3a and 4a of the main manuscript, the dI/dV characteristics for the d+id pairing potential (see left panel of Supplementary Fig. \ref{fig:did_long}) can be divided into two parts for this in-plane setup.

 The inner disk-like structure in Supplementary Fig. \ref{fig:graphene_long}a, limited by the minimal bandgap ($E = 6.8\Delta$), corresponds to the inner sunflower structure of the d+id pairing potential shown in Figs.~3a, 4a of the main manuscript. Here, again the Andreev reflections take place in the pseudo quantum well and the quasiparticles are confined if their energies are less than the amplitudes of both pair potentials, this is $E < \text{min} (\abs{\Delta_+}, \abs{\Delta_-})$. Again, the peak position moves between 0 and $\text{min}(\abs{\Delta_+}, \abs{\Delta_-})$, depending on the relation of $\Delta_+$, $\Delta_-$ and the angle $\alpha$. Furthermore, as in  Figs.~3a and 4a of the main manuscript, the local bandgap peak occurs if the energy E of the injected particle is $E = \abs{\Delta_{\pm}}$. As before, this finding is in analogy to the direct mapping of the maximum value of the pairing potential in certain $\alpha$-directions, as discussed for the short-range d+id case in the main text.

 For the f-wave pairing at the same doping $x=0.15$, we chose $U_0 = 10\mathrm{eV}$, $U_1 =5\mathrm{eV}$ and $U_2 = 3\mathrm{eV}$ (see Eq.~(5) in Sect. IV of the main text and Supplementary Fig.~ \ref{fig:graphene_long}b).
 Again our general statement holds, which is that the physics of the time-reversal-symmetry-preserving f-pairing case, for the longer-range Coulomb situation, is crucially different from the chiral d+id case: in the f-wave scenario, we only find the outer local bandgap peaks (rotated by an angle $\frac{\pi}{6}$ in respect to the pairing potential) plus the zero energy bound state peaks.

\section*{Supplementary Discussion 2: phenomenological differential conductance spectrum for cobaltates}
\begin{figure}[p]
\centering
\begin{overpic}[width=0.5 \linewidth]
{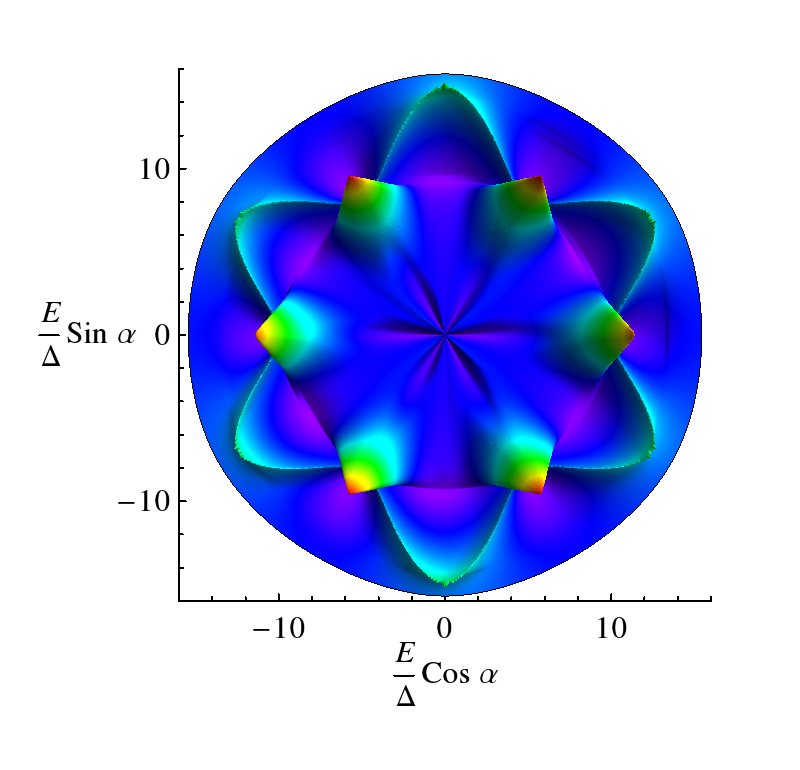}
\put(105,10){\includegraphics[width=0.07\linewidth]{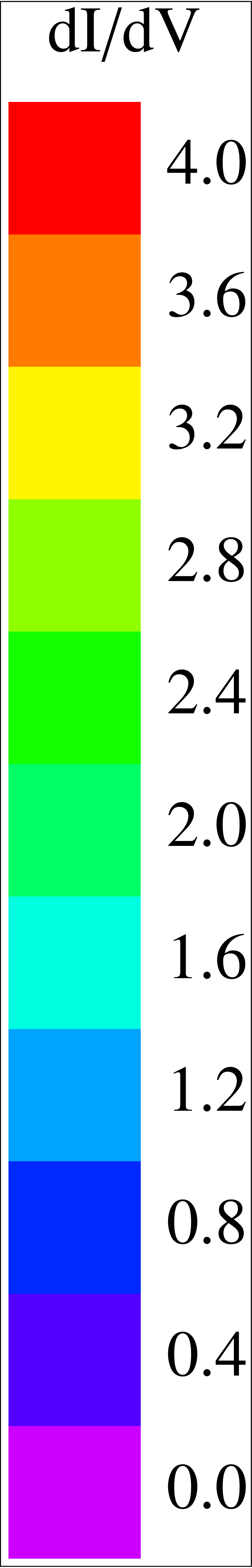}}
\end{overpic}
\caption{"Phenomenological" differential conductance spectrum (dI/dV) for the d+id superconducting phase of the cobaltates, if only the first harmonic is taken for the real and the imaginary part. The left panel shows the differential conductance spectrum, obtained in the in-plane setup simulating an STM experiment, plotted as a function of the quasiparticle energy E (radial axis, normalised by the reference band gap $\Delta$) and the angle $\alpha$ between the interface normal and the $k_{\mathrm{x}}$-direction of the pairing potential (polar axis). Differences in brightness of the colors are due to the plot style.}
\label{fig:didlessharm}
\end{figure}
Here we compare the full fRG + STM results for the d+id order parameter for cobaltates with a phenomenological STM approach: In this approach, one takes only the first harmonic of the superconducting (SC) gap function into account, which corresponds to assuming only short-range (nearest-neighbour) electron-electron 
interactions entering the pairing function. The corresponding differential conductance plot is shown in Supplementary Fig.~2 (for the in-plane setup). It is much simpler and qualitatively different from the one obtained in our main manuscript (see Fig.~4a) 
for the realistic dI/dV calculations. In the full fRG + STM calculations not only nearest-neighbour pairing interactions, but also next nearest neighbour and higher order interactions (higher harmonics) are important and give rise to a quite complex structure of the d+id order parameter.
More specifically, comparing Supplementary Fig.~2 with Fig.~4a of the main manuscript, one sees first, that the entire broken time reversal symmetry (BTRS) structure for small quasiparticle energies (in the center of the plot) is absent for the phenomenological order parameter. 
Second, the BTRS structure is shifted to the edge of the superconducting gap in Supplementary Fig. \ref{fig:didlessharm}. 
 
Third, the ratio of the intensities of the main peaks (ratio of BTRS structure vs. superconducting bandgap peak) is significantly changed.
Therefore, including the higher harmonics, obtained within the fRG procedure, provides qualitatively new insight into the differential conductance plot and is the necessary condition to unambiguously identify topological superconductivity in experiments.

\section*{Supplementary Discussion 3: differential conductance at finite temperature for cobaltates}
\begin{figure}[p]
\centering
\begin{overpic}[width=0.49 \linewidth]{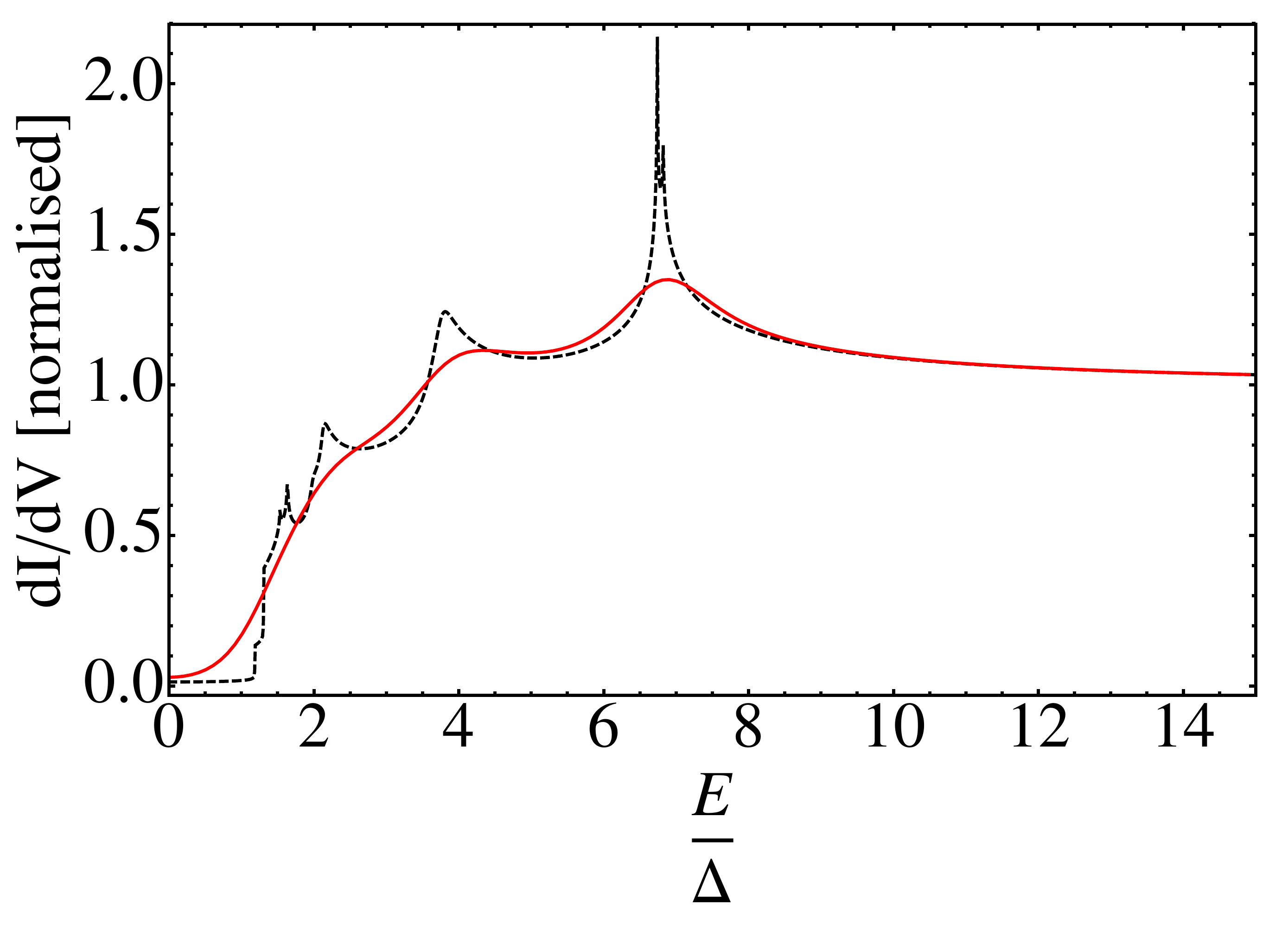}
\put(20,55){\includegraphics[width=25mm]{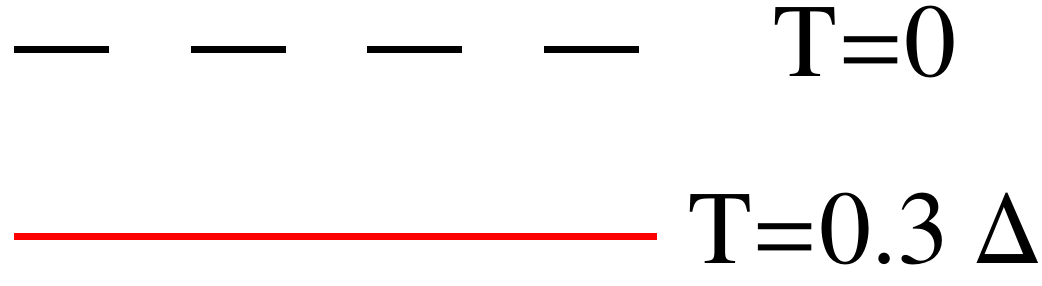}}\end{overpic}
\begin{overpic}[width=0.49 \linewidth]{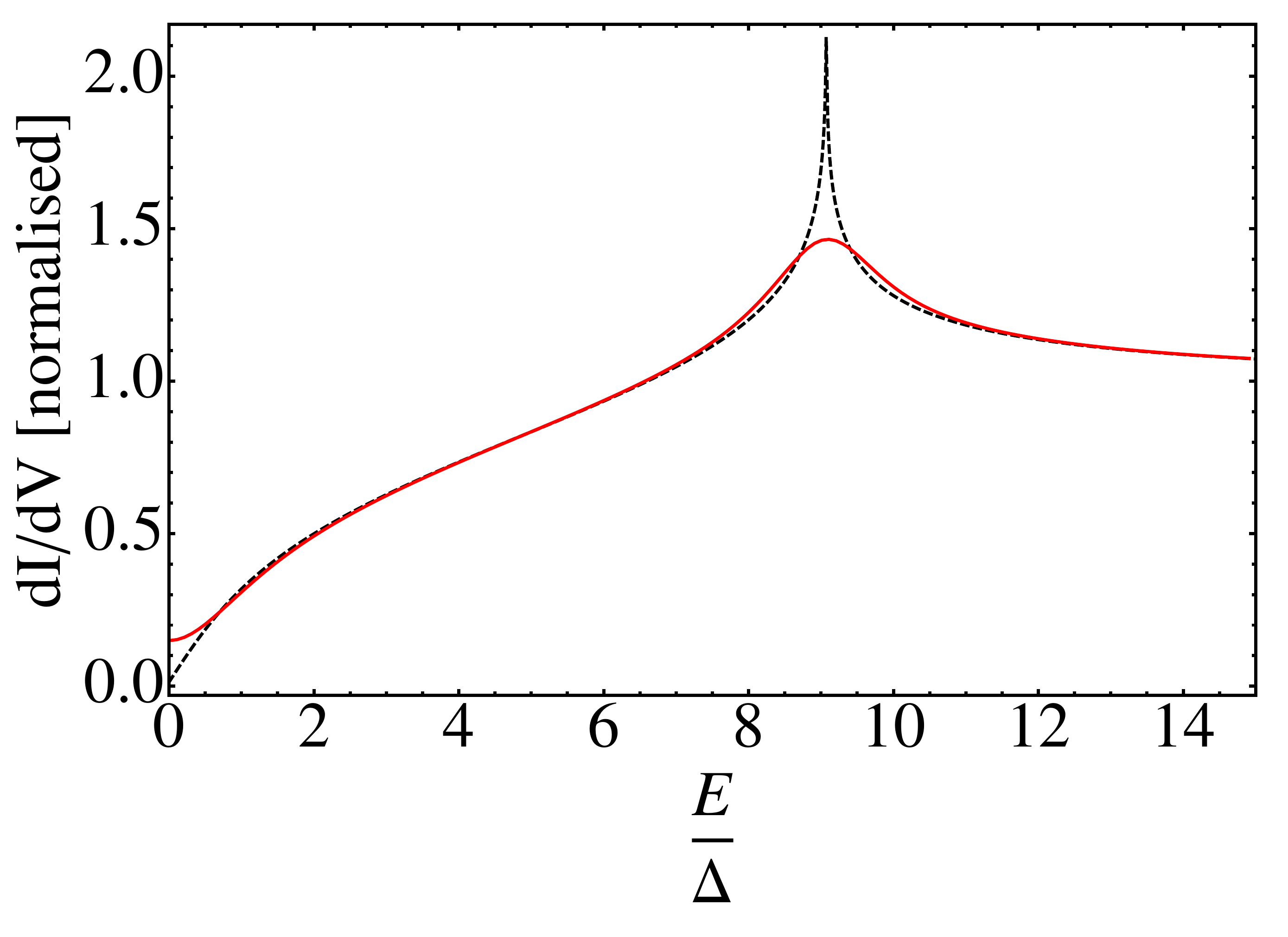}
\put(20,55){\includegraphics[width=25mm]{legend}}\end{overpic} \\
\begin{overpic}[width=0.49 \linewidth]{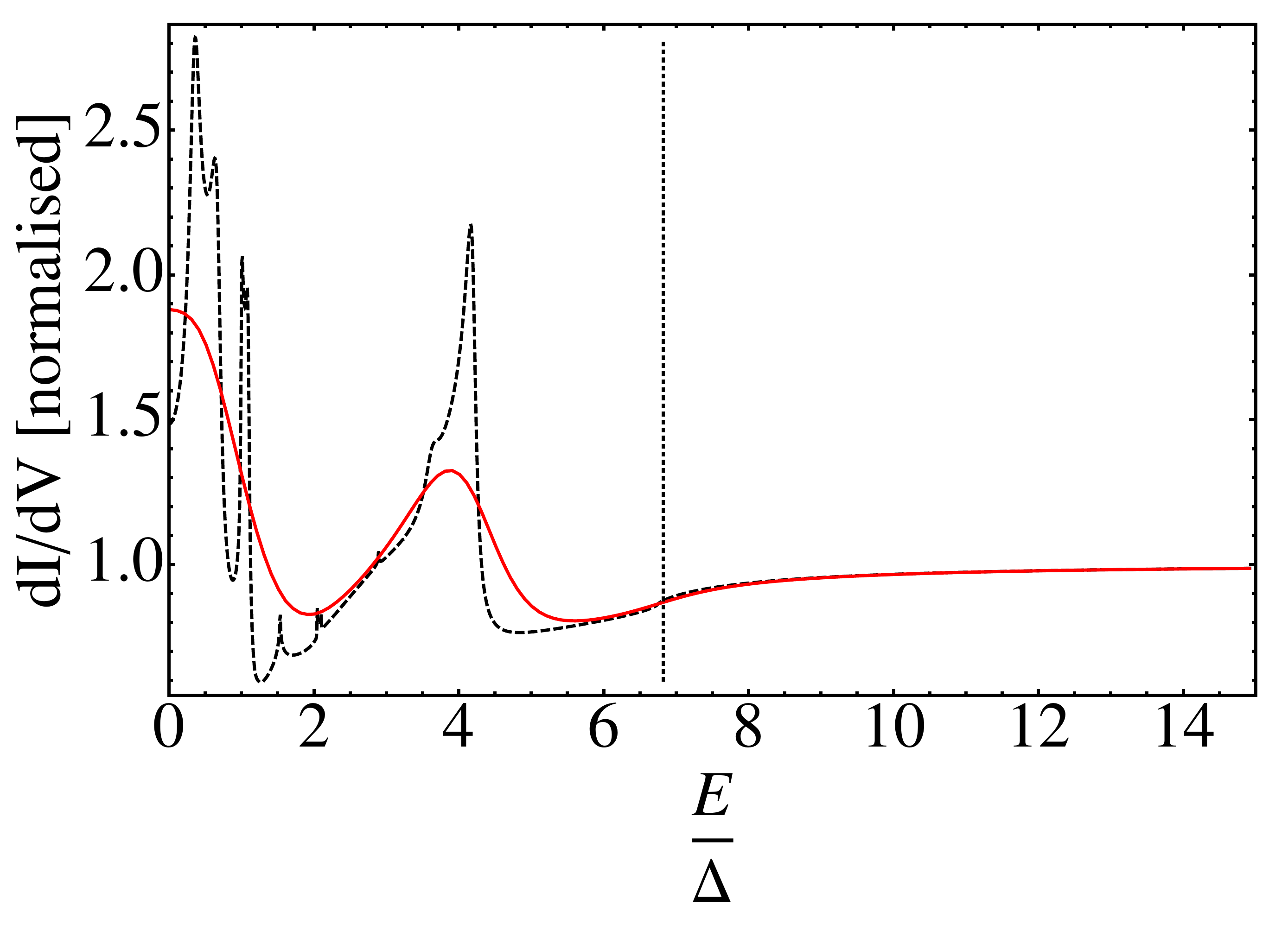}
\put(60,55){\includegraphics[width=25mm]{legend}}\end{overpic}
\begin{overpic}[width=0.49 \linewidth]{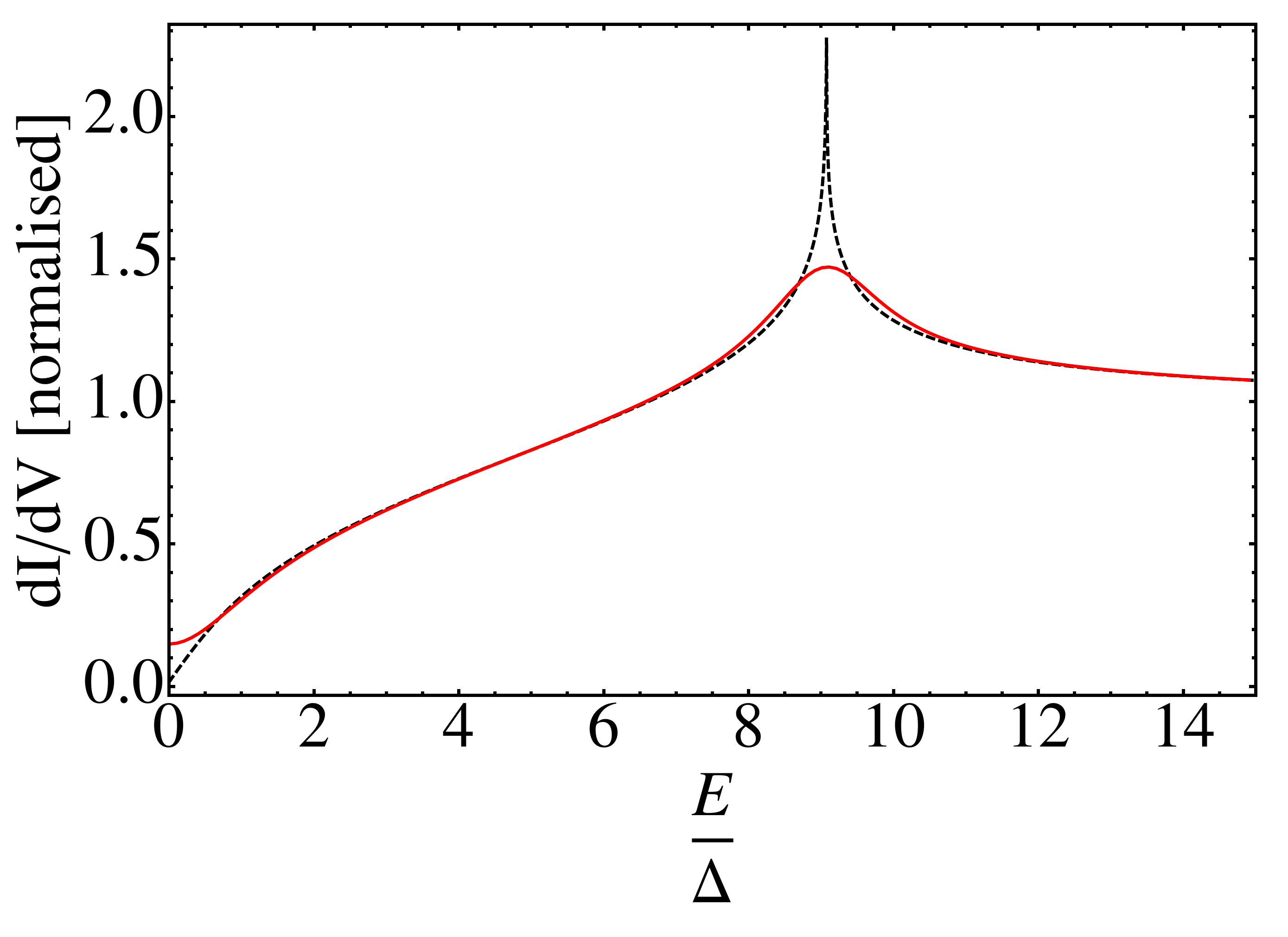}
\put(20,55){\includegraphics[width=25mm]{legend}}\end{overpic}
\caption{dI/dV characteristics for d+id and f order parameter of the cobaltates for in-plane and out-of-plane STM setups. The red curves show a small temperature broadening $k_{\mathrm{B}} T=0.3 \Delta$. $\alpha$ denotes the angle between the interface normal and the $k_{\mathrm{x}}$-direction. Upper left panel: d+id phase, out-of-plane setup. Upper right panel: f phase, out-of-plane setup. Lower left panel: d+id phase, in-plane setup, $\alpha=0$, the dotted line shows the maximum absolute value of the gap. Lower right panel: f phase, in-plane setup, $\alpha=0$.}
\label{fig:T}
\end{figure}
Supplementary Figs.~3a and 3b show the differential conductance for the d+id and f pairing phases of the cobaltates in the out-of-plane setup for zero temperature and for a finite temperature of $k_{\mathrm{B}} T = 0.3 \Delta$. Supplementary Figs.~3c and 3d give the corresponding plots in the in-plane setup (for a cut at $\alpha=0$). Note, that $\Delta \approx 1\mathrm{meV} \approx 12\mathrm{K}$.

The superconducting gap for the anisotropic d+id order parameter for the cobaltates is so small, that the weak temperature broadening closes the gap. Therefore, for small quasiparticle energies the d+id and f pairing phases cannot be distinguished in the differential conductance for the out-of plane STM setup. 
In contrast, as shown in Figs.~3c and 3d, the in-plane setup allows to clearly distinguish the two phases as shown for the $\alpha =0$ plane. While the differential conductance for the f-pairing shows just the local density of states peak (the signal looks identical to an out-of-plane geometry), 
the STM signal for the d+id SC pairing shows clear signatures of Andreev bound states which survive at finite temperature.

\section*{Supplementary Methods 1: harmonics for the pairing potential of graphene with short range Coulomb interactions}
The following four harmonics were used to fit the gap for the d+id pairing phase of graphene with short range Coulomb interactions. 
\begin{align}
d_{\mathrm{x}^2-\mathrm{y}^2}^{(1)} &= 
 2 \cos(\sqrt{3} k_{\mathrm{y}}) -\cos \left(\frac{\sqrt{3}k_{\mathrm{y}}-3k_{\mathrm{x}}}{2}\right) -\cos\left(\frac{\sqrt{3}k_{\mathrm{y}} + 3k_{\mathrm{x}}}{2}\right) \\
d_{\mathrm{xy}}^{(1)}    &= \sqrt{3}
 \cos\left(\frac{\sqrt{3}k_{\mathrm{y}} - 3k_{\mathrm{x}}}{2}\right) - \sqrt{3} \cos\left(\frac{\sqrt{3}k_{\mathrm{y}} +3k_{\mathrm{x}}}{2}\right)
\\
d_{\mathrm{x}^2-\mathrm{y}^2}^{(2)} &= 
2\cos(3k_{\mathrm{x}}) - \cos\left(\frac{3\sqrt{3}k_{\mathrm{y}}-3k_{\mathrm{x}}}{2}\right)-\cos\left(\frac{3\sqrt{3}k_{\mathrm{y}}+3k_{\mathrm{x}}}{2}\right)\\
d_{\mathrm{xy}}^{(2)} &= \sqrt{3}
 \cos\left(\frac{3\sqrt{3}k_{\mathrm{y}}-3k_{\mathrm{x}}}{2}\right)-\sqrt{3} \cos\left(\frac{3\sqrt{3}k_{\mathrm{y}}+3k_{\mathrm{x}}}{2}\right) 
\end{align}
Fitting these basis functions to the numerical data gives an analytical expression for the pairing potential.
\begin{eqnarray}
\Delta_{\mathrm{d+id}}^{\mathrm{short}} = \Delta \cdot (c_1  d_{\mathrm{x}^2-\mathrm{y}^2}^{(1)} + c_2 i d_{\mathrm{xy}}^{(1)} +c_3 d_{\mathrm{x}^2-\mathrm{y}^2}^{(2)} +c_4 i d_{\mathrm{xy}}^{(2)})
\end{eqnarray}
The fitting factors are $c_1=c_2=1.68$ and $c_3=c_4=0.32$.
$\Delta$ is an arbitrarily chosen superconducting energy scale, which is the same for all phases and allows to compare the energy scale of the phases. It is used for the normalisation of the energy in the plots. The same harmonics were used to obtain the expression for the long-range Coulomb interaction pairing phase.
For the cobaltates, 12 harmonics were used to achieve a fit.

\section*{Supplementary Methods 2: normalised differential conductance}

The wave functions for the normal part (N) and the superconducting part (S) are obtained solving the Bogoliubov-de Gennes Hamiltonian.
\begin{align}
\Psi_{\mathrm{N}}(r)&=e^{ik_{\mathrm{FN}}r}\zweivector{1}{0} + r_{\mathrm{A}} e^{ik_{\mathrm{FN}}r}\zweivector{0}{1}+r_{\mathrm{N}} e^{-ik_{\mathrm{FN}}r}\zweivector{1}{0}\\
\Psi_{\mathrm{S}}(r)&=t e^{ik_{\mathrm{FS}}r}\zweivector{u_+}{e^{-i\Phi_+}v_+}+t' e^{-ik_{\mathrm{FS}}r}\zweivector{e^{i\Phi_-}v_-}{u_-}
\end{align}
$u_{\pm}=\sqrt{\frac{E+\sqrt{E^2-\absq{\Delta_{\pm}}}}{2E}}$, $v_{\pm}=\sqrt{\frac{E-\sqrt{E^2-\absq{\Delta_{\pm}}}}{2E}}$ and $\Phi_{\pm}=\argument{\Delta_{\pm}}$. $\Delta_+$ ($\Delta_-$) denotes the pairing potential exhibited by the electron-like (hole-like) quasiparticles (ELQ/HLQ), where we define $\Delta_{\pm} = \Delta(\phi_{\pm})$. In the in-plane setup $\phi_+=\theta_{\mathrm{S}} -\alpha$ and $\phi_-=\pi-\theta_{\mathrm{S}} - \alpha$. $\theta_{\mathrm{S}}$ denotes the angle of an ELQ with respect to the interface normal and $\alpha$ gives the mismatch between the interface normal and the $k_{\mathrm{x}}$-direction of the pairing potential (see also Fig. 2, main part). If the Fermi momenta of the normal region and the superconductor are identical, $k_{\mathrm{FS}}=k_{\mathrm{FN}}$, then $\theta_{\mathrm{S}}=\theta_{\mathrm{N}}$=$\theta$. In the out-of plane setup $\phi_+ = \phi_- = \phi$, $\theta_+ = \theta_{\mathrm{S}}$ and $\theta_- = \pi -\theta_{\mathrm{S}}$, where the $\pm$ index refers to ELQ and HLQ, respectively. The insulator is modelled with a delta-Dirac barrier with effective, incident angle dependent, barrier height $Z = \frac{Z_0}{cos\theta}$, where $Z_0=\frac{mH}{\hbar^2 k_{\mathrm{FN}}}$.

 We use the continuity of the wave function at the interface and the matching condition of the derivative of the wave functions of the normal and the superconducting side to obtain the coefficients for Andreev ($r_{\mathrm{A}}$) and normal ($r_{\mathrm{N}}$) reflection.
The total conductance of the NIS junction is obtained by integrating the BTK conductance over all independent contributions, which is integrating over all $k_{\mathrm{y}}$ in our in-plane setup. We normalise the conductance by the conductance of a NIN junction in the same geometrical setup. The normalised differential conductance in the in-plane setup $\sigma^{\mathrm{in}}$ is
 \begin{equation}
\sigma^{\mathrm{in}} = \frac{\int \sigma_{\mathrm{S}}(E,\theta)\cos\theta d\theta}{\int \sigma_{\mathrm{N}}(E,\theta)\cos\theta d\theta}.
\end{equation}
In the out-of-plane setup, we have to integrate over all momenta lying in the plane of the superconductor. The normalised differential conductance in the out-of-plane setup $\sigma^{\mathrm{out}}$ is
\begin{equation}
\sigma^{\mathrm{out}} = \frac{\int_0^{\frac{\pi}{2}} \int_0^{2\pi} \sigma_{\mathrm{S}} (E, \theta, \phi) \sin\theta \cos\theta d\theta d\phi}{\int_0^{\frac{\pi}{2}} \int_0^{2\pi} \sigma_{\mathrm{N}} (E, \theta, \phi) \sin\theta \cos\theta d\theta d\phi}.
\end{equation}

\end{document}